\documentclass[a4paper,aps,showpacs,floatfix,twocolumn,%
superscriptaddress,pre]{revtex4-1}

\usepackage{amsmath}
\usepackage{color}
\usepackage{graphicx}
\usepackage{bbold}

\newcommand{\vect}[1]{\boldsymbol{#1}}

\def \eps{\varepsilon}
\def \wt{\widetilde}
\def \a{\alpha}
\def \b{\beta}
\def \p{\partial}
\def \g{\gamma}

\begin{document}

\title{Consistent two-phase Lattice Boltzmann model
for gas-liquid systems}

\author{Jasna Zelko}
\affiliation{Max Planck Institute for Polymer Research,
Ackermannweg 10, 55128 Mainz, Germany}
\author{Burkhard D\"unweg}
\affiliation{Max Planck Institute for Polymer Research,
Ackermannweg 10, 55128 Mainz, Germany}
\affiliation{Condensed Matter Physics, TU Darmstadt,
Karolinenplatz 5, 64289 Darmstadt, Germany}
\affiliation{Department of Chemical Engineering,
Monash University, Clayton, Victoria 3800, Australia}

\date{\today}

\begin{abstract}
  A new lattice Boltzmann method for simulating multiphase flows is
  developed theoretically. The method is adjusted such that its
  continuum limit is the Navier-Stokes equation, with a driving force
  derived from the Cahn-Hilliard free energy. In contrast to previous
  work, however, the bulk and interface terms are decoupled, the
  former being incorporated into the model through the local
  equilibrium populations, and the latter through a forcing term. We
  focus on gas-liquid phase equilibria with the possibility to
  implement an arbitrary equation of state. The most novel aspect of
  our approach is a systematic Chapman-Enskog expansion up to the
  third order. Due to the third-order gradient in the interface
  forcing term, this is needed for full consistency with both
  hydrodynamics and thermodynamics. Our construction of a model that
  satisfies all conditions is based upon previous work by Chen,
  Goldhirsch, and Orszag (J. Sci. Comp. 34, 87 (2008)), and implies 59
  and 21 velocities in three and two dimensions, respectively.
  Applying the conditions of positivity of weights, existence of a
  two-phase region in the phase diagram, and positivity of the bulk
  viscosity, we find substantial restrictions on the permitted
  equation of state, which can only be lifted by an even more refined
  model. Moreover, it turns out that it is necessary to solve a
  self-consistent equation for the hydrodynamic flow velocity, in
  order to enforce the identity of momentum density and mass current
  on the lattice. The analysis completely identifies all spurious
  terms in the Navier-Stokes equation, and thus shows how to
  systematically eliminate each of them, by constructing a suitable
  collision operator. The commonly noticed inconsistency of most
  existing models is thus traced back to their insufficient number of
  degrees of freedom. Therefore, the gain of the new model is in its
  clear derivation, full thermo-hydrodynamic consistency, and expected
  complete elimination of spurious currents in the continuum limit.
  Numerical tests are deferred to future work.
\end{abstract}

\pacs{
47.11.-j, 
47.55.Ca, 
05.20.Dd  
}

\maketitle

\section{\label{sec:intro}
Introduction}

The Lattice Boltzmann (LB) method \cite{succi_lattice_2001,
  benzi_lattice_1992,chen_lattice_1998,dunweg_lattice_2009,
  he_priori_1997,qian_lattice_1992}
is based upon solving a fully discretized version of the Boltzmann
equation known from the kinetic theory of gases. Space is discretized
in terms of the sites of a regular lattice with spacing $a$ and time
in terms of a finite time step $h$, while the velocity space is
reduced to a small set of discrete velocities $\vect c_i$ that are
chosen such that one time step will always connect sites on the
lattice (i.~e. $\vect c_i h$ is a lattice vector). The central objects
of the algorithm are the populations $n_i(\vect r, t)$ at site $\vect
r$ at time $t$, corresponding to the velocity $\vect c_i$, and the
algorithm proceeds by an alteration of streaming and collision
steps. Typically, $n_i$ is assigned the physical interpretation of a
mass density.

The simplest version of this algorithm is for an isothermal ideal gas,
where the collision step is done by linearly relaxing the populations
towards a set of local pseudo-equilibrium populations $n_i^{eq}$,
which in turn are determined from the local mass density $\rho =
\sum_i n_i$ and the local flow velocity $\vect u = \rho^{-1} \vect j =
\rho^{-1} \sum_i n_i \vect c_i$. This scheme can be analyzed in detail
by a multiple time-scale Chapman-Enskog (CE) expansion (see,
e.~g. Ref. \cite{dunweg_lattice_2009}), from which one finds that the
algorithm provides a valid solution to the isothermal Navier-Stokes
equation (NSE) in the continuum limit, provided that the Mach number
(flow velocity $u$ relative to the speed of sound $c_s$) is
sufficiently small, i.~e. terms of order $u^3$ may be safely
neglected. In turn, this means that the scheme is confined to flows
that are (close to) incompressible. It is this analysis that provides
essentially all of the deeper insights: one starts by introducing a
dimensionless scale separation parameter $\eps \ll 1$ and setting
$\vect r_1 = \eps \vect r$. At fixed $\vect r_1$, the limit $\eps \to
0$ then automatically implies an analysis at large length
scales. Since in standard hydrodynamics two time scales are involved,
one for sound waves, where $\text{time} \propto (\text{length})^1$,
and a slower one for diffusive momentum transfer, where $\text{time}
\propto (\text{length})^{2}$, the analysis takes this into account
by explicitly introducing two time variables, $t_1 = \eps t$
(wave-like scaling) and $t_2 = \eps^2 t$ (diffusive scaling), and
formally treating the dynamic variables as depending on $t_1$ and
$t_2$ independently. The limiting behavior is then obtained by a
leading-order Taylor expansion with respect to $\eps$. It should be
noted that the CE expansion may also be viewed as an expansion with
respect to gradients --- wave-like scaling corresponds to first-order
gradients, while diffusive terms imply second-order gradients in the
NSE. For the ideal-gas LB algorithm, the CE analysis then provides a
host of important results \cite{dunweg_lattice_2009}):
\begin{itemize}
\item The non-dissipative Euler equation is obtained at first order of
  the CE expansion, while dissipation corresponds to the second order.
\item The Euler dynamics is completely encoded in the algebraic form
  of the equilibrium populations, i.~e. in the dependence of
  $n_i^{eq}$ on $\rho$ and $\vect u$ --- and the analysis also shows
  that $n_i^{eq}$ may only depend on the conserved quantities, which
  are mass and momentum.
\item In order to avoid spurious terms in the Euler equation, one
  needs at least three velocity shells. In this case, the equation of
  state is fixed to $p = \rho c_s^2$, where $p$ is the thermodynamic
  pressure, while $c_s$ cannot be chosen at will, but takes the value
  $c_s = 3^{-1/2}$ in lattice units for the D3Q19 model
  \cite{qian_lattice_1992}. If another value is desired, more shells
  are needed.
\item Viscous dissipation corresponds to the relaxation towards local
  equilibrium, and the analysis provides explicit expressions for the
  relation between the shear and bulk viscosities in the NSE on the
  one hand, and the LB relaxation rates on the other.
\end{itemize}

Since roughly two decades, there has been extensive work that tries to
extend this algorithm to the case of multiphase flows, where the
interest was mainly focused on the case of a binary mixture on the one
hand, and the case of gas-liquid phase coexistence in a one-component
system on the other. The literature on this topic is vast, and hence
we do not attempt here to provide anything like a comprehensive
review, but rather only briefly mention the seminal work by Shan and
Chen \cite{shan_lattice_1993,shan_simulation_1994}, Swift and Yeomans
\cite{swift_lattice_1995,swift_lattice_1996}, and Lee and Fischer
\cite{lee_eliminating_2006}. Nevertheless, it seems that up to today
no fully consistent scheme has so far been constructed
\cite{nourgaliev_lattice_2003} --- meaning that it should be fully
compatible with thermodynamics, Galilean invariance, and free of
spurious currents. Although progress has been made in refining the
models and in reducing some of the artifacts
\cite{lee_eliminating_2006,
  he_discrete_1998,luo_unified_1998,he_thermodynamic_2002,wagner_origin_2003,
  cristea_reduction_2003,sofonea_finite-difference_2004,
  wagner_thermodynamic_2006,
  shan_analysis_2006,wagner_investigation_2006,sbragaglia_generalized_2007,
  pooley_eliminating_2008,guo_force_2011,lou_effects_2012,
  liu_three-dimensional_2012,liu_phase-field-based_2013}, we
nevertheless have the impression that a comprehensive solution of the
problem has so far been lacking.

In view of this unsatisfactory situation we here try to re-examine the
LB method for multiphase flows. In the present paper, we focus on the
conceptually simplest case, the isothermal gas-liquid system, which is
characterized by just two equations of motion, the mass conservation
and momentum conservation equations --- just as it is the case for the
isothermal ideal gas. Inspired by the success of the LB method for
this system, we try to construct the algorithm by making as much use
as possible of the lessons that we have learnt from
there. Nevertheless, there are very important differences, and it
seems that previous work did either not sufficiently appreciate these,
or did not systematically work out all the consequences, which, as it
unfortunately turns out, result in a fairly complicated and cumbersome
analysis. The main considerations that form the basis of our approach 
are the following:
\begin{itemize}
\item The concept of low Mach number flow needs to be scrutinized. The
  condition for small Mach number is $u^2 \ll \p p / \p \rho$, but the
  right-hand side \emph{vanishes} at the gas--liquid critical point,
  meaning that the condition \emph{cannot} hold under such
  circumstances. Below the critical point, it will hold in the pure
  liquid and gas phases, respectively; however these phases are
  connected by interfaces in which the density takes values within the
  coexistence region, where $\p p / \p \rho$ is small or even
  negative. On the other hand, we notice that in the ideal gas case
  the parameter $c_s^2$ can and should be considered as just the ratio
  $p / \rho$ (which here happens to coincide with $\p p / \p
  \rho$). Indeed, $c_s^2$ occurs in the equilibrium populations
  $n_i^{eq}$, and these encode the Euler stress, i.~e. the pressure as
  such and not $\p p / \p \rho$. It is then most natural to transfer
  this observation to the case of a non-ideal gas, i.~e. to assume
  that the non-ideal gas is characterized by an equation of state $p =
  \rho c_s^2(\rho)$ with a non-trivial density dependence of $c_s^2$,
  and, furthermore, to assume that the equilibrium populations have
  the \emph{same} form as in the ideal-gas case, only with a modified
  $c_s^2$.
\item This means that consistency with thermodynamics and Galilean
  invariance will necessarily involve more than three velocity shells.
\item We also notice that with this interpretation it is possible to
  achieve $u^2 \ll c_s^2$ throughout the phase diagram, which means
  that it should be justified to neglect $O(u^3)$ terms, just as in
  the CE analysis of the ideal gas.
\item Furthermore, the ideal-gas case tells us that we should view
  $n_i^{eq}$ as the solution to a maximum-entropy problem
  \cite{wagner_h-theorem_1998,karlin_perfect_1999,
    boghosian_galilean-invariant_2003,dunweg_statistical_2007}.
  Transferring this to the non-ideal gas, we find that the
  thermodynamics should probably best be viewed as derived from a
  non-ideal entropy, i.~e. a Boltzmann entropy with suitably adjusted
  non-trivial weights, which result in a non-trivial density
  dependence of the phase space volume in velocity space. We expect
  that this will ultimately pave the way towards a consistent
  formulation of the stochastic version of the method, which includes
  thermal fluctuations \cite{dunweg_statistical_2007}. This notion
  should not come as a big surprise to physicists with a background in
  modern soft-matter theory, where the concepts of energy and entropy
  are often used interchangably (what ultimately matters is just the
  free energy).
\item Aiming at thermodynamic consistency, it is then most natural to
  derive the macroscopic equations of motion from a Ginzburg-Landau
  type free energy functional, as in the LB approaches pioneered by
  Swift and Yeomans \cite{swift_lattice_1995,swift_lattice_1996}.
  Such a functional describes the bulk thermodynamics by a free energy
  density $f(\rho)$, which, below the critical point, exhibits a
  double-well structure, while phase separation is driven by an
  interface free energy, which is typically modeled by a
  gradient-square term in the functional.
\item Since $n_i^{eq}$ corresponds to the leading order of the CE
  expansion, which does not contain any gradients, $n_i^{eq}$ should
  correspond to \emph{only} the bulk free energy, or the bulk equation
  of state (and we have already outlined how to do this). In contrast,
  it should \emph{not} depend on the interfacial term, as it was
  introduced in the original Swift-Yeomans model
  \cite{swift_lattice_1995,swift_lattice_1996}.
\item Rather, we model the interface term similar to the effect of an
  external force. This is somewhat reminiscent of the work by Lee and
  Fischer \cite{lee_eliminating_2006}; however, they confined
  the driving not to the interface term (as we do), but rather to all
  terms that deviate from the ideal gas (which, in our opinion, is
  inconsistent as well).
\item On the NSE level, the interfacial driving shows up via a term
  that involves a \emph{third-order} gradient of the density. This
  means however that the CE analysis has to be done up to third order
  as well, which is obviously a cumbersome task. 
\item One is therefore naturally led to the introduction of a third time
  scale, $t_3 = \eps^3 t$, which is yet slower than momentum diffusion.
  We believe that the physical interpretation of the process corresponding
  to $t_3$ is simply domain coarsening, which is typically the slowest
  process in a phase-separating system.
\item Apart from the obvious terms in the collision operator
  (relaxation towards $n_i^{eq}$, interfacial force-like driving), we
  also construct a ``correction'' collision operator that cancels
  various spurious terms, such that the final equation of motion up to
  order $\eps^3$ is just the NSE. This correction is analogous to the
  correction that is known in the case of driving an ideal gas via an
  external force \cite{dunweg_lattice_2009,guo_discrete_2002}. It is
  essentially impossible to guess the form of such a correction
  without doing the CE analysis, and it is (we believe) a non-trivial
  result that it is possible to construct such an operator at all. We
  believe that this approach is most likely the only way to
  systematically eliminate all numerical artifacts from the method. A
  numerical test is however deferred to future work.
\end{itemize}

It should be noted that in previous work
\cite{wagner_thermodynamic_2006} it has already been realized that
multiphase LB models should be subjected to a higher-order CE
analysis. However, it seems that the present paper is the first
attempt in which this program is actually being carried out. The
analysis will show that the higher-order CE expansion also implies
more stringent requirements on the isotropy of lattice tensors. While
for the ideal gas (second-order CE expansion) isotropy up to
fourth-rank tensors is needed, the present model requires isotropy up
to sixth-rank tensors (and a concomitant larger set of velocities).
The fact that improved isotropy is helpful to construct better
multiphase LB models has been noted before as well
\cite{shan_analysis_2006, sbragaglia_generalized_2007}. Therefore, we
could resort to previous work on isotropy of lattice tensors; the
present paper builds upon results found by Chen, Goldhirsch and Orszag
\cite{chen_discrete_2008}. As an alternative, we might also have used
the formalism developed by Shan \cite{shan_general_2010}. We emphasize
that our solution is just one possible way (out of many) to construct
such a model, and probably not even the most efficient one.

The remainder of this paper is organized as follows: Section
\ref{sec:basic} outlines the general LB setup of our model, and
discusses the target NSE. In Sec. \ref{sec:velocity_c} we then discuss
how to find a proper set of velocities, and how to implement the
equation of state in terms of the equilibrium populations. Section
\ref{sec:centraldifference} then presents a brief excursion on
central-difference approximations to various gradients that need to be
evaluated on the lattice. Section \ref{sec:MomentumTransfer} discusses
how the momentum transfer derived from the interface force is split up
between the various contributions of the collision operator, and from
this we construct the interface force collision operator in
Sec. \ref{sec:interface}. The central part of the paper is then
Sec. \ref{sec:CE}, in which the CE analysis is done. The results
derived there then allow us to construct the correction collision
operator, which is done in Sec. \ref{sec:correction}.  Section
\ref{sec:implicit} outlines the implicit algorithm that needs to be
applied in order to find the hydrodynamic flow velocity with
third-order CE accuracy. Finally, we summarize in
Sec. \ref{sec:conclus}. Appendix \ref{app:modelH} derives the
third-order interface force from the Cahn-Hilliard model, while
App. \ref{app:EquationOfState} works out how to construct an equation
of state that is compatible with the various restrictions derived in
the main text. Some algebraic details that have been omitted in the
main text are presented in App. \ref{app:hardwork}.

\section{\label{sec:basic}
Basic considerations}

Our starting point are the hydrodynamic equations of motion that
should be simulated by the multiphase LB method. Mass conservation is
expressed by the continuity equation
\begin{equation} \label{eq:Continuity_intro}
  \p_t \rho + \p_\a \left( \rho u_\a \right) = 0 ,
\end{equation}
where $\rho(\vect r, t)$ is the mass density, $\vect u (\vect r, t)$
the flow velocity, Greek letters denote Cartesian indexes, the
Einstein summation convention is implied, and $\p_t \equiv \p / \p t$,
$\p_\a \equiv \p / \p r_\a$. Momentum conservation implies an equation of
motion for the momentum density $\vect j = \rho \vect u$, of the form
\begin{equation} \label{eq:NS}
  \p_t(\rho u_\a) + \p_\b (\rho u_\a u_\b) + \p_\a p 
  = \p_\b \sigma_{\a \b} + f_\a , 
\end{equation}
where $p$ is the thermodynamic pressure, $\vect f$ the interfacial
force density and $\sigma$ the viscous stress tensor involving the
shear viscosity $\eta$ and the bulk viscosity $\eta_V$:
\begin{equation}
  \sigma_{\a \b} = \eta \left[ \p_\a u_\b + \p_\b u_\a -
  \frac{2}{d} \p_\g u_\g \delta_{\a \b} \right]
  + \eta_V \p_\g u_\g \delta_{\a \b}  , 
  \label{eq:StressTensor}
\end{equation}
where $\delta_{\a \b}$ denotes the Kronecker symbol and $d$ the spatial
dimension. Pressure and interface forces are related to the
Cahn-Hilliard Hamiltonian \cite{cahn_free_1958}
\begin{equation} 
  {\cal H} = \int d^d \vect r \left[ \frac{1}{2} \rho
    \vect u^2 + \rho e + \frac{\kappa}{2} \left( \nabla \rho \right)^2
  \right] ,
\end{equation}
where the first term denotes the kinetic energy density, $e = e(\rho)$
is the internal energy per unit mass and $\kappa$ the interfacial
stiffness. In the absence of viscous dissipation, the dynamics should
conserve the total Hamiltonian,
\begin{equation}
  \frac{d}{dt} {\cal H} = 0 ,
\end{equation}
and since the pressure is related to $e$ via
\begin{equation}
  p = \rho^2 \frac{\p e}{\p \rho} ,
\end{equation}
this condition allows the determination of $\vect f$ (see
App. \ref{app:modelH}):
\begin{equation} \label{eq:ResultingForce}
  \vect f = \kappa \rho \nabla \nabla^2 \rho .
\end{equation}

LB simulations are based on the kinetic theory of gases and the
Boltzmann equation. The algorithm can be summarized by the following
discretized version of the Boltzmann equation:
\begin{equation} \label{eq:boltzmann}
  n_i(\vect r + \vect c_i h,t+h) = n_i^{*}(\vect r, t)
  = n_i(\vect r, t) + \Delta_i ,
\end{equation} 
where $n_i(\vect r, t)$ indicates the pre-collisional populations ---
mass density of particles at site $\vect r$ and at time $t$ that have
the velocity $\vect c_i$. $n_i^{*}$ indicates the post-collisional
populations. The difference between the pre- and post-collisional
populations $\Delta_i$ is the collision operator. The algorithm is
hence performed in two steps from the right to the left hand side; the
first step is the collision step and the second step is the streaming
step. In the streaming step the post-collisional populations $n_i^{*}$
are moved to neighboring sites $\vect r + \vect c_i h$, where $h$ is
the time step of the simulation. $\{\vect c_i h\}$ is therefore a
discrete set of vectors which connect each lattice site with its
neighbouring sites. The connection between this microscopic equation
and the macroscopic equations of motion
(Eqs. \ref{eq:Continuity_intro} and \ref{eq:NS}) is found via the CE
analysis. After Taylor expansion with respect to the scaling parameter
$\eps \ll 1$, one studies various velocity moments of the populations
and their equations of motion at different levels of the multiple
time-scale analysis. At the end the velocity moments of different
orders are gathered back together, so that the zeroth velocity moment
results in the continuity equation (Eq. \ref{eq:Continuity_intro}) and
the first velocity moment in the Navier-Stokes equation
(Eq. \ref{eq:NS}). As discussed in the Introduction, the third-order
gradient in the interface force requires that we introduce three time
scales for our CE analysis, and hence we write
\begin{align}
  \vect r_1 & = \eps \vect r \label{eq:r1},\\
  t_1      & = \eps t,\\
  t_2      & = \eps^2 t,\\
  t_3      & = \eps^3 t,
\end{align}
such that the corresponding space and time derivatives can be written as
\begin{align}
  \p_\a & = \eps \p_{\a_1} ,
  \label{eq:part_r}\\
  \p_t  & = \eps \p_{t_1} + \eps^2 \p_{t_2} + \eps^3 \p_{t_3} .
  \label{eq:part_t}
\end{align}
The time scales $t_1$ and $t_2$ are already known from the ideal gas
CE expansion \cite{dunweg_lattice_2009} and correspond to sound waves
and diffusion of momentum, respectively, while the physical
interpretation of the newly introduced time scale $t_3$ can be found
in the coarsening process of the multiphase system. With this the
lattice Boltzmann equation (Eq. \ref{eq:boltzmann}) is rewritten as
\begin{eqnarray}
  \nonumber
  &&
  n_i(\vect r_1 + \eps \vect c_i h, t_1 + \eps h, t_2 + \eps^2 h,
  t_3 + \eps^3 h)
  \\
  & - & 
  n_i(\vect r_1,t_1,t_2,t_3) = \Delta_i ,
  \label{eq:LatticeBoltzmann}
\end{eqnarray}
and its CE analysis will be done in Sec.~\ref{sec:CE}.

The collision operator $\Delta_i$ is composed of the bulk, the
interface and the correction term:
\begin{equation}
  \Delta_i = \Delta_i^{bulk} + \Delta_i^{int} + \Delta_i^{corr} \, . 
  \label{eq:CollisionOperator}
\end{equation}
For the sake of simplicity we just use the standard BGK
(Bhatnagar-Gross-Krook) operator for $\Delta_i^{bulk}$:
\begin{equation}
  \Delta_i^{bulk} = (\gamma - 1) (n_i - n_i^{eq}) ,
\end{equation}
where $\gamma$ with $-1 < \gamma < 1$ is the relaxation parameter. The
generalization to an MRT (multi relaxation time) collision operator is
in principle straightforward; however, this would, in view of the many
velocity shells that are needed, require additional somewhat
cumbersome algebra, which is deferred to future work. 

The interface collision operator $\Delta_i^{int}$ is constructed so
that the first velocity moment of $\Delta_i$ corresponds to the
interface force (Eq. \ref{eq:ResultingForce}). By this we assure that
the interface force is correctly implemented into the algorithm. The
detailed derivation is done in Secs. \ref{sec:MomentumTransfer} and
\ref{sec:interface}. The correction collision operator is constructed
so that all spurious terms, which would occur in the continuum
equations of motion in case there was no such correction operator, are
cancelled out exactly up to the third order in the CE expansion. The
final construction of the correction collision operator is done after
the CE expansion based on the same principle as previously used to
implement an external force in the ideal gas LB algorithm
\cite{guo_discrete_2002,dunweg_lattice_2009} (see
Sec. \ref{sec:correction}).

It turns out that the construction of the correction collision
operator is quite elaborate and requires an LB algorithm which is
isotropic up to sixth rank tensors, in contrast to standard algorithms
which usually satisfy the isotropy up to the fourth rank only. Because
of this a large number of velocities $\vect c_i$ is necessary. The
relation between lattice velocity sets and the tensors of their
moments has been worked out in great detail in previous papers
\cite{chen_discrete_2008,shan_general_2010}, and we were fortunate that
we could build on that existing body of work. Our concrete
implementation is based upon Ref. \cite{chen_discrete_2008}. Within
that formalism, we found that on the two-dimensional square lattice 21
velocities are needed to satisfy isotropy up the sixth rank, and at
the same time still keep the freedom to pick $c_s^2$ at will, in order
to obtain a proper non-ideal equation of state:
\begin{equation}
  p(\rho) = \rho c_s^2(\rho)  . \label{eq:ofState}
\end{equation}
In three dimensions, analogous considerations (again within the same
formalism) yield an even larger set of 59 velocities on the
simple-cubic lattice. In what follows, we will restrict attention to
these two lattices.

A further constraint is the requirement that the weights $w_i$ that
are assigned to each velocity shell in the construction of $n_i^{eq}$
(see Sec. \ref{sec:velocity_c}) must be strictly positive. This is
because in the entropic \cite{wagner_h-theorem_1998,
  boghosian_galilean-invariant_2003,karlin_perfect_1999} and the
stochastic \cite{dunweg_statistical_2007} generalization of the method
the $w_i$ take directly the role of statistical weights or
probabilities. In other words, to violate that condition would prevent
the possibility to construct a consistent thermodynamics or
statistical mechanics of the model, and most likely fundamental
problems with the validity of the second law (or, actually, the mere
possibility to \emph{define} an entropy) would arise. In practice,
this means that $c_s^2$ is allowed to vary only within narrow bounds
set by the velocity set, and hence the density ratio between the gas
and the liquid phase is quite limited as well.

We thus see that in order to achieve a consistent multiphase LB model
some of the advantages of the ideal gas LB model have to be
sacrificed. We already mentioned the need of a significantly larger
number of velocity vectors. In fact, most of the existing models try
to stick to the same number of velocities as needed for the ideal gas,
which is, of course, an additional source of inconsistency. Besides
this, because of the interface force and correction collision
operators, which are based on the evaluation of gradients of
thermodynamic variables, the collision step is no more localized at a
lattice site, but has to involve neighboring sites. This is in fact
common to most if not all of the existing LB multiphase models. One
further complication arises in our algorithm. In order to derive an
exact continuity equation up to the third order, a redefinition of the
momentum density is needed, which implies an implicit algorithm for
the calculation of the flow velocity. This is derived in
Sec. \ref{sec:CE} and elaborated further in Sec. \ref{sec:implicit}.

\section{\label{sec:velocity_c}
Velocity set and equilibrium populations}

The equilibrium populations $n_i^{eq}$ may be viewed as the discrete
version of the Maxwell-Boltzmann distribution function. We define them
here exactly as in the case of an ideal gas, with the only difference
that we allow for a density dependence of both $c_s^2$ and of the
weights $w_i$:
\begin{eqnarray}
  \label{eq:equilibrium}
  & &
  n_i^{eq} (\rho, \vect u)
  \\
  \nonumber
  & = & 
  w_i (\rho) \rho 
  \left( 1 + \frac{\vect u \cdot \vect c_i}{c_s^2(\rho)} 
  + \frac{(\vect u \cdot \vect c_i)^2}{2 c_s^4(\rho)} -
  \frac{u^2}{2 c_s^2(\rho)} \right) .
\end{eqnarray}
In practice, the evaluation of $n_i^{eq}$ has to be done as follows:
(i) Firstly, one determines $\rho = \sum_i n_i$, and then (ii) reads
off the value of $c_s^2$ from the equation of state (for details on
this, see App. \ref{app:EquationOfState}). This permits (iii) the
determination of all the weights $w_i$ (see below). Furthermore, one
needs to (iv) evaluate the interface force density $\vect f$ (see
Sec. \ref{sec:interface}), as well as a (v) ``correction current''
$\tilde{\vect j}$ (this is explained within the framework of the CE
analysis, see Secs. \ref{sec:CE} and \ref{sec:implicit}), and then
(vi) determine the flow velocity $\vect u$ via the prescription (see
also Sec. \ref{sec:MomentumTransfer})
\begin{equation}
  \label{eq:DefineJCurrent}
  \vect j = \rho \vect u = \sum_i n_i \vect c_i + 
  \frac{h}{2} \vect f + \tilde{\vect j} .
\end{equation}

We first discuss the set of velocities and weights that we use in our
model. It is clear that for symmetry reasons the weights must be
identical for vectors within the same shell, i.~e. vectors that have
the same length. Lattice tensors composed from the velocities $\vect
c_i$ via moments weighted with the $w_i$ have to satisfy sixth rank
isotropy (this is needed for the proper construction of the correction
collision operator as described in Sec. \ref{sec:correction}). Odd
moments trivially vanish for reasons of reflection symmetry of the
lattice. For the even orders cubic symmetry and proper normalization
implies
\begin{align}
  & \sum_i w_i = 1 , 
    \label{eq:tensor0} \\
  & \sum_i w_i c_{i\a} c_{i\b} = \sigma_2 \delta_{\a\b},
    \label{eq:tensor2}\\
  & \sum_i w_i c_{i\a} c_{i\b} c_{i\g} c_{i\delta} = \kappa_4 \delta_{\a\b\g\delta}
    \label{eq:tensor4}
  \\
  \notag
  & + \sigma_4 (\delta_{\a\b}\delta_{\g\delta} + \text{perm.}),
  \\
  & \sum_i w_i c_{i\a} c_{i\b} c_{i\g} c_{i\delta} c_{i\mu} c_{i\nu} = 
    \kappa_6 \delta_{\a\b\g\delta\mu\nu}
  \label{eq:tensor6}
  \\
  \notag
  & + \eta_6 (\delta_{\a\b}\delta_{\g\delta\mu\nu} + \text{perm.}) 
    + \sigma_6(\delta_{\a\b}\delta_{\g\delta}\delta_{\mu\nu} + \text{perm.}) .
\end{align}
Here the various $\delta$ symbols are one if all indexes are identical
and zero otherwise. Furthermore, the term ``perm.'' means that one has
to take into account all possibilities to assign subsets of indexes to
the various $\delta$ symbols, such that, for example, the bracket near
$\sigma_4$ contains three terms and the bracket near $\sigma_6$
contains 15 terms.

The isotropy condition requires that $\kappa_4 = 0$, $\kappa_6 = 0$,
$\eta_6 = 0$. Furthermore, we require $\sigma_4 = \sigma_2^2$ as well
as $\sigma_6 = \sigma_2^3$. These latter two conditions can be
motivated from the continuum analog, i.~e. the Maxwell-Boltzmann
velocity distribution at vanishing flow velocity, $\vect u = 0$. This
is a Gaussian distribution, whose tensorial moments result from Wick's
theorem, and have the same form as given above. We also require,
in analogy to the ideal-gas case, that $c_s^2 = \sigma_2$. The
moment relations are therefore simplified to
\begin{align}
  & \sum_i w_i = 1 , 
    \label{eq:tensor0_revised}
  \\
  & \sum_i w_i c_{i\a} c_{i\b} = c_s^2 \delta_{\a\b} ,
    \label{eq:tensor2_revised}
  \\
  & \sum_i w_i c_{i\a} c_{i\b} c_{i\g} c_{i\delta} = 
    c_s^4 ( \delta_{\a\b} \delta_{\g\delta} + \text{perm.} ) ,
  \label{eq:tensor4_revised}
  \\
  & \sum_i w_i c_{i\a} c_{i\b} c_{i\g} c_{i\delta} c_{i\mu} c_{i\nu} = 
    c_s^6 ( \delta_{\a\b} \delta_{\g\delta} \delta_{\mu\nu} 
    + \text{perm.} ) .
  \label{eq:tensor6_revised}
\end{align}
For a freely picked value of $c_s^2$, we thus have six conditions for
the weights $w_i$ ($\kappa_4 = 0$, $\kappa_6 = 0$, $\eta_6 = 0$,
$\sigma_4 = c_s^4$, $\sigma_6 = c_s^6$, and normalization), resulting
in seven ``degrees of freedom'' or velocity shells that are expected
to be needed in order to satisfy all constraints. As a matter of fact,
however, we use eight shells in three dimensions, while in two
dimensions it is possible to satisfy all conditions with only six
shells. The set of shells and the corresponding weights are given in
Tabs. \ref{tab:vel3D} and \ref{tab:vel2D} for the 3D and 2D cases,
respectively. They have been constructed by making use of the work of
Chen et al. \cite{chen_discrete_2008}, and we refer interested readers
to that paper. Briefly, the D3Q59 model is obtained by projecting a
set of 4D lattice velocities which satisfy sixth rank isotropy down to
3D, while the D2Q21 model is obtained by further projecting the D3Q59
model down to two dimensions. In particular, the weights for the
lower-dimensional sets are directly obtained from the weights of the
original 4D model.

\begin{table}
\begin{tabular}{|l |c| l|}
  \hline 
  $b$ & 
  $\vect c_i$ 
  & weight \\ 
  \hline 
  & & \\
  1 & (000) & 
  $w_0 = 1 - \frac{2351}{720} \sigma_2 + \frac{1081}{192} \sigma_2^2 
  - \frac{363}{192} \sigma_2^3$ 
  \\[15pt]
  6 & (100) & 
  $w_1 = 2 \sigma_2 (\frac{8}{45} - \frac{1}{3}\sigma_2 
  + \frac{1}{6} \sigma_2^2)$
 \\[15pt]
  12 & (110) &
 $w_2 = \sigma_2 (\frac{8}{45} - \frac{1}{3} \sigma_2
 + \frac{1}{6} \sigma_2^2)$
 \\[15pt]
 8 & (111) & 
 $w_3 = \frac{1}{12} \sigma_2 (-\frac{16}{15} + 4 \sigma_2
 - \frac{5}{2} \sigma_2^2)$
 \\[15pt]
 6 & (200) &
 $w_4 = \frac{1}{96} \sigma_2 (-\frac{24}{5} + 17 \sigma_2 - 9 \sigma_2^2)$
 \\[15pt]
 12 & (220) & 
 $w_5 = \frac{1}{192} \sigma_2 (-\frac{8}{15} + \sigma_2 + \sigma_2^2)$
 \\[15pt]
 8 & (222) & 
 $w_6 = \frac{1}{192} \sigma_2 (\frac{4}{15} - \sigma_2 + \sigma_2^2)$
 \\[15pt]
 6 & (400) & 
 $w_7 = \frac{1}{384} \sigma_2(\frac{4}{15} - \sigma_2 + \sigma_2^2)$
 \\ 
 \hline
\end{tabular}
\caption{Eight velocity shells and corresponding weights needed
  to construct a 3D model. The total number of velocities is 59.
  $b$ is the number of velocities within one shell. For each
  shell, we only list one representative lattice vector.
  We here use lattice units, i.~e. $h = a = 1$.}
\label{tab:vel3D}
\end{table}

\begin{table}
\begin{tabular}{|l |c| l|}
  \hline 
  $b$ & 
  $\vect c_i$ & 
  weight \\ 
  \hline 
  & &\\
  1 & (00) & 
  $w_0 = 1 - \frac{45}{2} \sigma_2 (\frac{7}{60} - \frac{7}{48} \sigma_2
  + \frac{1}{16} \sigma_2^2)$
  \\[15pt]
  4 & (10) &
  $w_1 = \frac{1}{3} \sigma_2 (\frac{32}{15} - 4 \sigma_2 + 2 \sigma_2^2)$
  \\[15pt]
  4 & (11) & $w_2 = \sigma_2 (\frac{1}{3} \sigma_2 - \frac{1}{4} \sigma_2^2)$
  \\[15pt]
  4 & (20) & 
  $w_3 = \sigma_2 (-\frac{1}{18} + \frac{3}{16} \sigma_2
  - \frac{1}{12}\sigma_2^2)$
  \\[15pt]
  4 & (22) &
  $w_4 = \frac{1}{96} \sigma_2 (-\frac{1}{2} \sigma_2 
  + \frac{3}{2} \sigma_2^2)$
  \\[15pt]
  4 & (40) &
  $w_5 = \frac{1}{384} \sigma_2 (\frac{4}{15} - \sigma_2 + \sigma_2^2)$
  \\
  \hline
\end{tabular}
\caption{Six velocity shells and corresponding weights needed
  to construct a 2D model. The total number of velocities is 21.
  $b$ is the number of velocities within one shell. For each
  shell, we only list one representative lattice vector.
  We here use lattice units, i.~e. $h = a = 1$.}
\label{tab:vel2D}
\end{table}

With these conditions on the moments of the $w_i$, we can evaluate
moments of $n_i^{eq}$. As in the ideal-gas case, the low-order moments
are the hydrodynamic variables mass density, momentum density, and
Euler stress:
\begin{align}
  & \sum_i n_i^{eq} = \rho , \label{eq:mass00} \\
  & \sum_i n_i^{eq} c_{i\a} = j_\a = \rho u_\a , \label{eq:momentum00} \\
  & \sum_i n_i^{eq} c_{i\a} c_{i\b} = \pi_{\a\b}^{eq}
    = \rho c_s^2(\rho) \delta_{\a\b} + \rho u_\a u_\b 
  \notag
  \\
  & = p(\rho) \delta_{\a\b} + \rho u_\a u_\b .
  \label{eq:euler00}
\end{align}
Furthermore, we evaluate the third and fourth order moments as
\begin{align}
  & \sum_i n_i^{eq} c_{i\a} c_{i\b} c_{i \g} =
  \phi_{\a\b\g}^{eq}
  \notag
  \\
  & = \rho c_s^2 (\delta_{\a\b} u_{\g} 
  + \delta_{\a\g} u_\b + \delta_{\b\g} u _\a) ,
  \label{eq:phi00}
  \\
  & \sum_i n_i^{eq} c_{i\a} c_{i\b} c_{i \g} c_{i \delta} =
  \psi_{\a\b\g\delta}^{eq} 
  \notag
  \\
  & = \rho c_s^4
  (\delta_{\a\b} \delta_{\g\delta} 
  + \delta_{\a\g} \delta_{\b\delta}
  + \delta_{\a\delta} \delta_{\b\g}) 
  \notag
  \\
  & + \rho c_s^2
  (\delta_{\a\b} u_\g u_\delta
  + \delta_{\a\g} u_\b u_\delta
  + \delta_{\a\delta} u_\b u_\g )
  \notag
  \\
  & + \rho c_s^2
  (\delta_{\g\delta}  u_\a u_\b 
  +\delta_{\b\delta}  u_\a u_\g 
  +  \delta_{\b\g} u_\a u_\delta) .
  \label{eq:psi00}
\end{align}

Since the weights have to be positive, $w_i > 0$, $c_s^2$ is limited
to a certain interval, i.~e. $c_{s, min}^2 < c_s^2 < c_{s, max}^2$. Graphical
analysis of the $w_i$ reveals (in lattice units) that $c_{s, min}^2$
takes the values 0.3510760 and 0.3850612 for the two- and three-dimensional
models, respectively, while the corresponding values for $c_{s, max}^2$ are
1.3333333 and 1.1917145. This means that one is limited in the
construction of the equation of state, too (cf. Eq. \ref{eq:ofState}).
More details on this are found in App. \ref{app:EquationOfState}.

\section{\label{sec:centraldifference}
Discretization of derivatives}

\subsection{Lattice sets and Taylor expansion}

The interface collision operator and the correction collision operator
will turn out to depend on different orders of derivatives of
hydrodynamic variables like density $\rho$ or velocity $\vect u$. For
the CE expansion and the implementation of the algorithm we hence need
to know how different orders of derivatives are systematically
discretized. This is explained within this section.

We start from an arbitrary but symmetric set of \emph{dimensionless}
lattice vectors $\vect d_i$ with dimensionless coefficients
$\tau_i$. Again we organize the vectors in shells and require that the
$\tau_i$ are the same within each shell. In contrast to the weights
$w_i$, there is no restriction on the sign of the $\tau_i$, and there
is in general no normalization condition either. The odd moments of
$\tau_i$ vanish for symmetry reasons, while the even moments are
defined similarly as in Eqs. \ref{eq:tensor0}--\ref{eq:tensor4} up to
fourth order:
\begin{align}
  & \sum_i \tau_i =  \wt \sigma_0 , \label{eq:tensor_d0}\\
  & \sum_i \tau_i d_{i\a} d_{i\b} = \wt \sigma_2 \delta_{\a\b}, 
  \label{eq:tensor_d2}\\
  & \sum_i \tau_i d_{i\a} d_{i\b} d_{i\g} d_{i\delta} = \wt \sigma_4 
  (\delta_{\a\b}\delta_{\g\delta} + \text{perm.}) ;
  \label{eq:tensor_d4}
\end{align}
again we require that the fourth moment is isotropic whenever it occurs.

We assume that $g = g(\vect r_1)$ is some hydrodynamic variable like
mass density $\rho$, momentum density $\vect j$, or flow velocity
$\vect u = \vect j / \rho$. Within the framework of the CE expansion
terms of the form
\begin{equation}
  g_i \equiv g (\vect r_1 + \eps a \vect d_i)
\end{equation}
will appear, and they should be expanded (at least up to third order)
with respect to $\eps$:
\begin{eqnarray} \label{eq:taylor}
g_i & = & g + 
\eps a d_{i \alpha} \partial_{\alpha_1} g +
\frac{1}{2} \eps^2 a^2 d_{i \alpha} d_{i \beta} 
\partial_{\alpha_1} \partial_{\beta_1} g \label{eq:force_discretize} \\
\nonumber
& + &
\frac{1}{6} \eps^3 a^3 d_{i \alpha} d_{i \beta} d_{i \gamma}
\partial_{\alpha_1} \partial_{\beta_1} \partial_{\gamma_1} g +
O(\eps^4) .
\end{eqnarray}

We will now discuss discretizations case by case. For each case, we
will derive a set of conditions that a specific set of vectors $\vect
d_i$ and its set of coefficients $\tau_i$ has to fulfill in order to
correctly calculate the respective derivative up to third
order. Different derivatives are constructed by taking different
tensorial moments of Eq. \ref{eq:force_discretize} and by using
Eqs. \ref{eq:tensor_d0}--\ref{eq:tensor_d4}.

\begin{table}
\begin{tabular}{|c|c| c|c|c|c|}
\hline 
No. of vect. & state $\vect d_i$ & 
$\p_\a$ & $\p_{\a}\p_{\a}$ & $\overline{\p_\a\p_\b}$ & 
$\p_\a\p_\b\p_\b$ \\ 
\hline \hline
\multicolumn{6}{|c|}{2D} \\ 
\hline
1 & (00) & 0 	 	& -4	& 0	& 0	  \\
4 & (10) & 2/3 	        & 1	& 1	& -2      \\
4 & (11) & 0		& 0	& 1/4	& 1/2     \\
4 & (20) & -1/24	& 0	& 0	& 1/4	  \\
 \hline 
\multicolumn{6}{|c|}{3D} \\ 
\hline
1  & (000) & 0 	        & -6	& 0	& 0	   \\
6  & (100) & 2/3 	& 1	& 1/2   & -3       \\
12 & (110) & 0	        & 0	& 1/4	& 1/2	   \\
6  & (200) & -1/24	& 0	& 0	& 1/4	   \\
\hline
\end{tabular}
\caption{Coefficients $\tau_i$ of lattice vectors $\vect d_i$ for 2D and 3D
  constructions of specific derivatives. In the first column the number of
  vectors in the shell is given and in the second column one representative
  vector.}
\label{tab:derivatives}
\end{table}

\subsection{\label{sec:A}First derivative $\partial_\alpha g$}

This is a vector, and therefore it is sufficient to study a vectorial
moment:
\begin{eqnarray}
& &
\nonumber
\sum_i \tau_i d_{i \alpha} g_i =
\eps a \wt\sigma_2  \partial_{\alpha_1} g \\
& & +
\frac{1}{2} \eps^3 a^3 \wt\sigma_4 
\partial_{\alpha_1} \partial_{\beta_1} \partial_{\beta_1} g
+ O(\eps^4) .
\end{eqnarray}
We now require
\begin{eqnarray}
\wt\sigma_2 & = & 1 , \label{eq:cond1a}\\
\wt\sigma_4 & = & 0 ,\label{eq:cond1b}
\end{eqnarray}
resulting in
\begin{equation}
\frac{1}{a} \sum_i \tau_i d_{i \alpha} g_i =
\eps \partial_{\alpha_1} g+ O(\eps^4) ,
\end{equation}
or, by using Eq. \ref{eq:part_r},
\begin{equation}
\frac{1}{a} \sum_i \tau_i d_{i \alpha} g_i =
\partial_{\alpha} g + O(\eps^4) , \label{eq:der1}
\end{equation}
i.~e. a valid approximation for the first derivative. The set of
vectors $\vect d_i$ and the corresponding coefficients $\tau_i$ which
satisfy Eqs. \ref{eq:cond1a} and \ref{eq:cond1b} are listed in
Tab. \ref{tab:derivatives}.

\subsection{\label{sec:laplacian} 
Second derivative $\partial_\alpha \partial_\alpha g$}

This is a scalar, and hence one should study a simple scalar moment:
\begin{equation}
\sum_i \tau_i g_i = \wt\sigma_0 g + \frac{1}{2} \eps^2 a^2
\wt\sigma_2 \partial_{\alpha_1} \partial_{\alpha_1} g + O(\eps^4) ,
\end{equation}
We now require
\begin{eqnarray}
\wt\sigma_0 & = & 0 , \\
\wt\sigma_2 & = & 2,
\end{eqnarray}
resulting in
\begin{equation}
\frac{1}{a^2} \sum_i \tau_i g_i = 
\partial_{\alpha} \partial_{\alpha} g + O(\eps^4) ,
\label{eq:der2a}
\end{equation}
i.~e. a valid approximation for the Laplacian. The relevant set of
vectors and the corresponding weights are listed in
Tab. \ref{tab:derivatives}.

\subsection{Second derivative $\partial_\alpha \partial_\beta g$}

This is a symmetric second--rank tensor, and its trace has already
been discussed in the previous subsection. We therefore confine
attention to its traceless part,
\begin{equation}
\overline{\partial_\alpha \partial_\beta} g =
\left( \partial_\alpha \partial_\beta - \frac{1}{d} \delta_{\alpha \beta}
\partial_\gamma \partial_\gamma \right) g ,
\end{equation}
where $d$ is the spatial dimension, and, correspondingly, study the
traceless moment
\begin{equation}
\sum_i \tau_i \overline{d_{i \alpha} d_{i \beta}} g_i =
\wt \sigma_4 a^2 \overline{\partial_\alpha \partial_\beta} g + O(\eps^4) ,
\end{equation}
where the traceless part of the tensor $d_{i \alpha} d_{i \beta}$
is given by
\begin{equation}
\overline{d_{i \alpha} d_{i \beta}} = d_{i \alpha} d_{i \beta} 
- \frac{1}{d} \delta_{\alpha \beta} d_{i \gamma} d_{i \gamma} .
\end{equation}
Requiring $\wt \sigma_4 = 1$ then results in
\begin{equation}
\frac{1}{a^2} \sum_i \tau_i \overline{d_{i \alpha} d_{i \beta}} g_i =
\overline{\partial_\alpha \partial_\beta} g + O(\eps^4) ,
\end{equation}
i.~e. a valid approximation for the desired derivative. The
corresponding coefficients are again given in
Tab. \ref{tab:derivatives}. For the non-traceless tensor
$\partial_\alpha \partial_\beta g$ one simply has to superimpose
the present result with the approximation derived in the previous
subsection.

\subsection{\label{sec:D}Third derivative $\partial_\alpha 
\partial_\beta \partial_\beta g$}

This is again a vector, so a first moment is sufficient. We find
\begin{equation}
\sum_i \tau_i d_{i \alpha} g_i =
a \wt\sigma_2  \partial_{\alpha} g +
\frac{1}{2} a^3 \wt\sigma_4 
\partial_{\alpha} \partial_{\beta} \partial_{\beta} g
+ O(\eps^4) .\label{eq:der3_start}
\end{equation}
We now require
\begin{eqnarray}
\wt\sigma_2 & = & 0, \label{eq:tauSigma2}\\
\wt\sigma_4 & = & 2, \label{eq:tauSigma4}
\end{eqnarray}
resulting in
\begin{equation}
\frac{1}{a^3} \sum_i \tau_i d_{i \alpha} g_i =
\partial_{\alpha} \partial_{\beta} \partial_{\beta} g
+ O(\eps^4) ,\label{eq:der3}
\end{equation}
i.~e. just the desired derivative. The coefficients are again given in
Tab. \ref{tab:derivatives}.

\section{\label{sec:MomentumTransfer}
Momentum transfer}

We start from the condition that the total momentum density is changed
as a result of the acting force:
\begin{eqnarray}
  & & h f_\a = \sum_i \Delta_i c_{i\a} = \\
  \nonumber
  & & \sum_i \Delta_i^{bulk} c_{i\a} + \sum_i \Delta_i^{int} c_{i\a} +
      \sum_i \Delta_i^{corr} c_{i\a} = \\
  \nonumber
  & & \sum_i \left(\g - 1 \right) ( n_i - n_i^{eq} ) c_{i\a} +
      \sum_i \Delta_i^{int} c_{i\a} + \sum_i \Delta_i^{corr} c_{i\a} .
\end{eqnarray}
Abbreviating
\begin{eqnarray}
  \sum_i n_i c_{i\a} & =: & j_{0\a} , \\
  \Delta_i^{int} + \Delta_i^{corr} & =: & \Delta_i^{\prime} ,
\end{eqnarray}
we thus find
\begin{equation} \label{eq:momentum_transfer_1}
h f_\a = (\g - 1) (j_{0\a} - j_\a) + \sum_i \Delta_i^{\prime} c_{i\a} .
\end{equation}
On the other hand, the momentum density $j_\a$ is defined via the
prescription
\begin{equation} \label{eq:definition_of_j_current}
j_\a = \rho u_\a = \frac{1}{2} \sum_i (n_i + n_i^*) c_{i\a}
+ \tilde j_\a ,
\end{equation}
where the first term involves the arithmetic mean of the pre- ($n_i$)
and post-collisional ($n_i^* = n_i + \Delta_i$) populations, while the
second is the correction current. We can therefore write
\begin{eqnarray}
\nonumber
j_\a & = & 
\sum_i n_i c_{i\a} + \frac{1}{2} \sum_i \Delta_i c_{i\a} + \tilde j_\a \\
& = & j_{0\a} + \frac{1}{2} h f_\a + \tilde j_\a .
\label{eq:definition_of_j_current_new}
\end{eqnarray}
Combining Eqs. \ref{eq:momentum_transfer_1} and
\ref{eq:definition_of_j_current_new}, we can eliminate $j_{0\a} -
j_\a$ to find
\begin{equation}
\sum_i \Delta_i^{\prime} c_{i\a} = \frac{1 + \gamma}{2} h f_\a
+ (\gamma - 1) \tilde j_\a .
\end{equation}
Since $\Delta_i^{\prime} = \Delta_i^{int} + \Delta_i^{corr}$, it is most
natural to require that
\begin{eqnarray}
\label{eq:InterfaceMomentumTransferAppendix}
\sum_i \Delta_i^{int} c_{i\a} & = & \frac{1 + \gamma}{2} h f_\a , \\
\sum_i \Delta_i^{corr} c_{i\a} & = & (\gamma - 1) \tilde j_\a ;
\label{eq:CorrectionMomentumTransfer}
\end{eqnarray}
note that $\Delta_i^{int}$ should be a result of the interface force,
while both $\Delta_i^{corr}$ and $\tilde j_\a$ are correction terms derived
within the framework of the CE analysis. Actually, the two terms correspond
to different CE orders --- $f_\a$ is of third order, while $\tilde j_\a$
is of second order.

For the momentum transfer of the BGK part we thus obtain
\begin{equation}
\sum_i \Delta_i^{bulk} c_{i\a} = \frac{1 - \gamma}{2} h f_\a 
+ (1 - \gamma) \tilde j_a .
\end{equation}

\section{\label{sec:interface}
Interface force collision operator}

On the continuum level, the interface force density is given by
(cf. Eq. \ref{eq:ResultingForce}):
\begin{equation}
  f_\a = \kappa \rho \p_\a \p_\b \p_\b \rho .
\end{equation}
According to the results of Sec. \ref{sec:D}, $f_\a$ can hence be
approximated on the lattice as
\begin{equation}
  f_\a = \frac{\kappa \rho}{a^3} \sum_i \tau_i d_{i \alpha} \rho_i .
\end{equation}
Furthermore, from Sec. \ref{sec:MomentumTransfer} we know that
the interface collision operator should be mass-conserving,
and have a first velocity moment
\begin{equation}
  \sum_i \Delta_i^{int} c_{i\a} = \frac{1 + \g}{2} h f_{\a} .
  \label{eq:moment1int}
\end{equation}
It is easy to show that these conditions are met by the operator
\begin{equation}
  \Delta_i^{int} = \frac{1 + \g}{2} h 
  \frac{w_i(\rho_0)}{c_s^2(\rho_0)} c_{i\a} f_{\a} ;
\end{equation}
here $\rho_0$ is some arbitrarily chosen reference density. The second
velocity moment of this operator is evidently zero, while its third
moment is easily evaluated as
\begin{eqnarray}
& & \sum_i \Delta_i^{int} c_{i\a} c_{i\b} c_{i\g} 
\\
\nonumber
& = &
\frac{1 + \g}{2} h c_s^2(\rho_0) \left(
f_\a \delta_{\b \g} +
f_\b \delta_{\a \g} +
f_\g \delta_{\a \b} \right) .
\end{eqnarray}
Since the evaluation of $f_\a$ involves a third-order derivative,
$f_\a$ is of third order in the CE expansion, and this is true as well
for $\Delta_i^{int}$ and all of its moments.

Finally, we note that the procedure does not conserve the momentum
on the single lattice site; nevertheless, the global momentum
is strictly conserved. This is so because of the relation
\begin{equation}
\sum_{\vect r} \vect f(\vect r) = \frac{\kappa}{a^3}
\sum_i \tau_i \sum_{\vect r}
\rho(\vect r) \vect d_i \rho (\vect r + a \vect d_i)
= 0 .
\end{equation}
The vanishing of the total force is due to the fact that in the
inner sum each pair of densities occurs twice, with weighting
vectors $+ \vect d$ and $- \vect d$, respectively. These terms
therefore exactly cancel. A prerequisite is however that the system is
translationally invariant, which is the case for periodic boundary
conditions.

\section{\label{sec:CE} Chapman-Enskog analysis}

\subsection{CE hierarchy}

After these preliminary considerations, we are prepared for the CE
analysis of the algorithm, which will allow us to derive
$\Delta_i^{corr}$. We start from the LB equation
(cf. Eq. \ref{eq:LatticeBoltzmann})
\begin{eqnarray}
  \nonumber
  & &
  n_i(\vect r_1 + \eps \vect c_i h, t_1 + \eps h, t_2 + \eps^2 h,
  t_3 + \eps^3 h) 
  \\
  & - & n_i(\vect r_1,t_1,t_2,t_3) = \Delta_i \, .
\end{eqnarray}
Introducing the differential operator
\begin{equation}
D_i = \eps h c_{i\a} \p_{\a_1} + \eps h \p_{t_1} + \eps^2 h \p_{t_2} 
+ \eps^3 h \p_{t_3} ,
\label{eq:D}
\end{equation}
the LB equation can be re-written exactly as
\begin{equation}
  \left( \exp(D_i) - 1 \right) n_i = \Delta_i,
\end{equation}
or
\begin{equation}
  D_i n_i = D_i \left( \exp(D_i) - 1 \right)^{-1} \Delta_i ,
\end{equation}
and the right hand side can be expanded as a series involving
the Bernoulli numbers:
\begin{equation}
  D_i n_i = \left[ 1 - \frac{D_i}{2} + \frac{D_i^2}{12} + \ldots \right]
  \Delta_i .
  \label{eq:dif1}
\end{equation}
Furthermore, $n_i$ and $\Delta_i$ are also expanded in terms of the
parameter $\eps$ up to third order:
\begin{align}
  & n_i = n_i^{(0)} + \eps n_i^{(1)} + \eps^2 n_i^{(2)} + \eps^3 n_i^{(3)}
   + \ldots , \label{eq:ni} \\
  & \Delta_i = \Delta_i^{(0)} + \eps \Delta_i^{(1)} + \eps^2 \Delta_i^{(2)}
   +\eps^3 \Delta_i^{(3)} + \ldots \quad . \label{eq:deltai}
\end{align}
Inserting Eqs. \ref{eq:ni}, \ref{eq:deltai} and \ref{eq:D} into
Eq. \ref{eq:dif1} we get a systematic expansion in $\eps$, which
has to be satisfied at each order separately:
\begin{itemize}
  \item $\eps^{0}$:
    \begin{equation}
      \Delta_i^{(0)} = 0 .
    \end{equation}
  \item $\eps^{1}$:
    \begin{equation}
      (c_{i\a} \p_{\a_1} + \p_{t_1}) n_i^{(0)} 
      = \frac{1}{h} \Delta_i^{(1)} .
      \label{eq:eps1}
    \end{equation}
  \item $\eps^{2}$:
    \begin{eqnarray}
      \nonumber
      & &
      (c_{i\a} \p_{\a_1} + \p_{t_1}) n_i^{(1)} + \p_{t_2} n_i^{(0)}
      \\
      & = &
      \frac{1}{h} \Delta_i^{(2)} - \frac{1}{2} 
      (c_{i\a}\p_{\a_1} + \p_{t_1}) \Delta_i^{(1)} .
      \label{eq:eps2}
    \end{eqnarray}
    \item $\eps^{3}$:
      \begin{eqnarray}
        \nonumber
        & &
        (c_{i\a}\p_{\a_1} + \p_{t_1}) n_i^{(2)} +
        \p_{t_2} n_i^{(1)} +
        \p_{t_3} n_i^{(0)}
        \\ 
        \nonumber
        & = &
        \frac{1}{h} \Delta_i^{(3)} 
        - \frac{1}{2} (c_{i\a}\p_{\a_1} + \p_{t_1}) \Delta_i^{(2)}
        \\
        & - &
        \frac{1}{2} \p_{t_2} \Delta_i^{(1)} + 
        \frac{1}{12} h (c_{i\a}\p_{\a_1} + \p_{t_1})^2 \Delta_i^{(1)} .
        \label{eq:eps3}
      \end{eqnarray}
\end{itemize}
Since both $\Delta_i^{int}$ and $\Delta_i^{corr}$ are of higher order,
the zeroth-order equation is only of importance for the BGK
operator. This however means that we can identify the equilibrium
populations with the zeroth order: $n_i^{(0)} \equiv n_i^{eq}$, and we
can use the notations ``$(0)$'' and ``$eq$'' interchangably, both for
the populations and their moments. The next step will involve taking
velocity moments of the CE hierarchy.

\subsection{\label{sec:momentsPopulations}
Velocity moments}

We hence define:
\begin{itemize}
  \item zeroth moment: mass density
    \begin{equation}
      \rho = \sum_i n_i .
      \label{eq:mass}
    \end{equation}
  \item first moment: momentum density
    \begin{equation}
      j_\a = \sum_i n_i c_{i\a} + \frac{h}{2} f_\a + \wt j_\a .
      \label{eq:momentum}
    \end{equation}
  \item second moment: stress
    \begin{equation}
      \pi_{\a\b} = \sum_i n_i c_{i\a} c_{i\b} .
    \end{equation}
  \item third moment:
    \begin{equation}
      \phi_{\a\b\g} = \sum_i n_i c_{i\a} c_{i\b} c_{i\g} .
    \end{equation}
  \item fourth moment:
    \begin{equation}
      \psi_{\a\b\g\delta} = \sum_i n_i c_{i\a} c_{i\b} c_{i\g} c_{i\delta} .
    \end{equation}
\end{itemize}
By replacing $n_i$ with $n_i^{(k)}$, we obtain the corresponding
moments at $k$th order of the CE expansion, such that, e.~g.,
$\pi_{\a\b}^{(1)}$ denotes the first-order stress, i.~e. the second
velocity moment of $n_i^{(1)}$. For $k = 0$ (equilibrium populations)
these moments have already been derived at the end of
Sec. \ref{sec:velocity_c}. For $\rho$ and $\vect j$ it should be noted
that these are the hydrodynamic variables for which there are no
higher-order contributions, i.~e.  $\rho^{(1)} = \rho^{(2)} = \ldots =
0$, $\vect j^{(1)} = \vect j^{(2)} = \ldots = 0$.

The momentum density in Eq. \ref{eq:momentum} is defined as a mean
value between pre- and post-collisional momentum density plus an
additional term: $\vect j = (1/2) \sum_i (n_i + n_i^*) \vect c_i +
\tilde{\vect j}$, see also Sec. \ref{sec:MomentumTransfer}. The
additional term $\tilde{\vect j}$ is needed to guarantee the
continuity equation to be consistent up to the third order. Because of
this an implicit algorithm for the calculation of the fluid velocity $
\vect u = \vect j / \rho$ is required. This will be discussed in more
detail later.

We can re-write Eq. \ref{eq:momentum} as
\begin{equation}
  \sum_i n_i c_{i\a} = j_\a - \frac{h}{2} f_\a - \wt j_\a
\end{equation}
and specify for each order (note that $\vect f$ is of third order,
$\vect f =\eps^3 {\vect f}^{(3)} +  O(\eps^4)$, and $\wt{\vect j}$ cannot have a
zeroth-order contribution):
\begin{align}
  & \sum_i n_i^{(0)} c_{i\a} = j_\a , \\
  & \sum_i n_i^{(1)} c_{i\a} = - \wt j_\a^{(1)} , \\
  & \sum_i n_i^{(2)} c_{i\a} = - \wt j_\a^{(2)} , \\
  & \sum_i n_i^{(3)} c_{i\a} = - \frac{h}{2} f_\a^{(3)} - \wt j_\a^{(3)} .
\end{align}
We will find later that $\wt{\vect j}^{(1)} = \wt{\vect j}^{(3)} = 0$,
i.~e. that $\wt{\vect j}$ is a pure second-order contribution.

The analogous relation for the zeroth moment is
\begin{align}
  & \sum_i n_i^{(0)} = \rho , \\
  & \sum_i n_i^{(k)} = 0 \quad \text{for $k \ge 1$} .
\end{align}

Similarly, we need to discuss moments of the collision operator. For
the stress we make use of the fact that $\Delta_i = n_i^* - n_i$, and
hence we can write
\begin{equation}
  \sum_i \Delta_i^{(k)} c_{i \a} c_{i \b} = 
  \pi_{\a\b}^{*(k)} - \pi_{\a\b}^{(k)} ,
\end{equation}
and we will make use of similar expressions for the higher-order
moments as well. For the zeroth moment we note the mass conservation
condition $\sum_i \Delta_i = 0$, and hence
\begin{equation}
\sum_i \Delta_i^{(k)} = 0 \quad \text{for all $k$} .
\end{equation}
For the first moment, we know the momentum-transfer condition
$\sum_i \Delta_i \vect c_i = h \vect f$, and that the rhs is
of third order. Hence
\begin{align}
  & \sum_i \Delta_i^{(k)} \vect c_i = 0 
    \quad \text{for $k = 0, 1, 2$} , \\
  & \sum_i \Delta_i^{(3)} \vect c_i = h \vect f^{(3)} .
\end{align}

\subsection{\label{sec:SumVsDifference}
Pre- and post-collisional moments}

For the later development, it will be useful to know some relations
between pre- and post-collisional moments. The collision operator is
\begin{equation}
\Delta_i = (\g - 1) (n_i - n_i^{eq}) + \Delta_i^{int} + \Delta_i^{corr} .
\end{equation}
Introducing the notation $n_i^{neq} = n_i - n_i^{eq}$ for the
non-equilibrium populations, the update rule is
\begin{equation}
n_i^{* neq} = \gamma n_i^{neq} + \Delta_i^{int} + \Delta_i^{corr} .
\end{equation}
In view of later results, it will be useful to re-write this as
\begin{eqnarray}
\nonumber
\frac{1}{2} \left( n_i^{* neq} + n_i^{neq} \right) & = &
\frac{1}{2} \frac{\g + 1}{\g - 1} \left( n_i^{* neq} - n_i^{neq} \right)
\\
& + & \frac{1}{1 - \gamma} \left( \Delta_i^{int} + \Delta_i^{corr} \right) .
\end{eqnarray}
We now consider the second and third velocity moment of this
relation, at first and second order of the CE expansion. Since
$\Delta_i^{int}$ is of third order, it does not contribute.
For the moments of $\Delta_i^{corr}$ we introduce the abbreviations
\begin{eqnarray}
  \sum_{i} \Delta_i^{corr} c_{i\a} c_{i\b} & =: & \Sigma_{\a\b} ,
  \label{eq:SecondMomentDelta}\\
  \sum_{i} \Delta_i^{corr} c_{i\a} c_{i\b} c_{i\g} & =: & \Xi_{\a\b\g} ,
  \label{eq:ThirdMomentDelta}
\end{eqnarray}
and hence we have
\begin{eqnarray}
  \nonumber
  \frac{1}{2} \left( \pi_{\a\b}^{*(1,2)} + \pi_{\a\b}^{(1,2)} \right)
  & = & \frac{1}{2} \frac{\g+1}{\g-1}
  \left( \pi_{\a\b}^{*(1,2)} - \pi_{\a\b}^{(1,2)} \right) \\
  & + & \frac{1}{1 - \g} \Sigma_{\a\b}^{(1,2)} ,
  \label{eq:plusminuspi}\\
  \nonumber
  \frac{1}{2} \left( \phi_{\a\b\g}^{*(1,2)} + \phi_{\a\b\g}^{(1,2)} \right)
  & = & \frac{1}{2} \frac{\g+1}{\g-1}
  \left( \phi_{\a\b\g}^{*(1,2)} - \phi_{\a\b\g}^{(1,2)} \right) \\
  & + & \frac{1}{1 - \g} \Xi_{\a\b\g}^{(1,2)} ,
  \label{eq:plusminusphi}
\end{eqnarray}
where $(1,2) $ means that either first or second order can be
expressed this way.

\subsection{Mass conservation equation}

By taking the zeroth velocity moment of the CE hierarchy,
we find
\begin{itemize}
  \item at order $\eps^{1}$:
    \begin{equation}
      \p_{t_1} \rho + \p_{\a_1} j_{\a} = 0 .
      \label{eq:0eps1}
    \end{equation}
  \item at order $\eps^{2}$:
    \begin{equation}
      \p_{t_2} \rho  - \p_{\a_1}\wt j_\a^{(1)} = 0 .
      \label{eq:0eps2}
    \end{equation}
  \item at order $\eps^{3}$:
    \begin{equation}
      \p_{t_3} \rho - \p_{\a_1}\wt j_\a^{(2)}
      = \frac{h}{12}
      \p_{\a_1} \p_{\b_1} \left( \pi_{\a\b}^{*(1)} - \pi_{\a\b}^{(1)} \right) .
      \label{eq:MassConservationThirdOrder}
    \end{equation}
\end{itemize}
At this point, it becomes clear why the correction current $\wt{\vect
  j}$ is needed --- its purpose is to compensate the rhs of
Eq. \ref{eq:MassConservationThirdOrder}. We therefore require
\begin{equation}
  \wt j_\a^{(1)} = 0 \quad \text{and} \quad  
  \wt j_\a^{(2)} = - \frac{h}{12} \p_{\b_1}
  \left( \pi_{\a\b}^{*(1)} - \pi_{\a\b}^{(1)} \right) .
  \label{eq:wtj}
\end{equation}
We then multiply each equation of motion with its corresponding power
of $\eps$ and add up the resulting equations. This gives rise to the
continuity equation
\begin{equation}
  \p_{t} \rho + \p_\a j_\a = 0 ,
\end{equation}
which is therefore, by construction, accurate up to third order. We
will later see that $\wt j_\a^{(3)}$ does not appear in the equations,
and therefore we may assume that this order vanishes. We will discuss
later how to actually determine $\wt{\vect j}$.

\subsection{Momentum conservation equation}

Taking the first velocity moment of the CE hierarchy,
we obtain
\begin{itemize}
  \item at order $\eps^{1}$:
    \begin{equation}
      \p_{t_1} j_\a + \p_{\b_1} \pi_{\a\b}^{(0)} = 0 .
      \label{eq:1eps1}
    \end{equation}
  \item at order $\eps^{2}$:
    \begin{equation}
      \p_{t_2}j_\a + \frac{1}{2} \p_{\b_1} 
      \left( \pi_{\a\b}^{*(1)} + \pi_{\a\b}^{(1)} \right) = 0 ;
      \label{eq:1eps2}
    \end{equation}
    note that here we have made use of $\wt{\vect j}^{(1)} = 0$.
  \item at order $\eps^{3}$:
    \begin{eqnarray}
      \nonumber
      & &
      \p_{t_3} j_\a + \frac{1}{2} \p_{\b_1} 
      \left( \pi_{\a\b}^{*(2)} + \pi_{\a\b}^{(2)} \right) = f_\a^{(3)} 
      \\
      \nonumber
      & + & 
      \frac{h}{12} \p_{\b_1} \p_{\g_1} 
      \left( \phi_{\a\b\g}^{*(1)} - \phi_{\a\b\g}^{(1)} \right)
      \\
      & + &
      \frac{h}{12} \p_{\b_1} \p_{t_1} 
      \left( \pi_{\a\b}^{*(1)} - \pi_{\a\b}^{(1)} \right) ;
      \label{eq:1eps3}
    \end{eqnarray}
    here we have made use of Eq. \ref{eq:wtj} for both the
    first and the second order of $\wt{\vect j}$.
\end{itemize}
Adding the different orders results in
\begin{align}
  & \p_t j_\a + \p_\b \pi_{\a\b}^{(0)} + \eps \frac{1}{2} \p_\b
  \left( \pi_{\a\b}^{*(1)} + \pi_{\a\b}^{(1)} \right) = f_\a
  \label{eq:CEPrelim} \\
  & - \eps^2 \frac{1}{2} \p_\b 
  \left( \pi_{\a\b}^{*(2)} + \pi_{\a\b}^{(2)} \right) 
  \notag
  \\
  & + \eps^2 \frac{h}{12} \p_\b 
  \left[ \p_{\g_1}  \left( \phi_{\a\b\g}^{*(1)} - \phi_{\a\b\g}^{(1)} \right)
    + \p_{t_1}  \left( \pi_{\a\b}^{*(1)} - \pi_{\a\b}^{(1)} \right) \right] .
  \notag
\end{align}
We now make use of the results derived in Sec. \ref{sec:SumVsDifference}
to re-write this as
\begin{align}
  & \p_t j_\a + \p_\b \pi_{\a\b}^{(0)} + \eps \frac{1}{2} \frac{\g+1}{\g-1}
    \p_\b \left( \pi_{\a\b}^{*(1)} - \pi_{\a\b}^{(1)} \right) = f_\a
  \label{eq:CE} \\
  & - \eps^2 \frac{1}{2} \frac{\g+1}{\g-1} \p_\b
  \left( \pi_{\a\b}^{*(2)} - \pi_{\a\b}^{(2)} \right) 
  \notag
  \\
  & + \eps^2 \frac{h}{12} \p_\b 
  \left[ \p_{\g_1}  \left( \phi_{\a\b\g}^{*(1)} - \phi_{\a\b\g}^{(1)} \right)
    + \p_{t_1}  \left( \pi_{\a\b}^{*(1)} - \pi_{\a\b}^{(1)} \right) \right]
  \notag
  \\
  & - \eps   \frac{1}{1 - \g} \p_\b \Sigma^{(1)}_{\a \b}
    - \eps^2 \frac{1}{1 - \g} \p_\b \Sigma^{(2)}_{\a \b} .
  \notag
\end{align}
This should finally be the Navier-Stokes (NS)
equation. $\pi_{\a\b}^{(0)}$ is the Euler stress
(cf. Eq. \ref{eq:euler00}), and the analysis will show that the term
on the lhs with the prefactor $\eps$ is the divergence of the
Newtonian viscous stress. All terms on the rhs except the interface
forcing term are spurious, and hence should cancel. It will turn out
that this is possible by a suitable adjustment of the correction
collision operator. In order to proceed, we need to close the
equation, i.~e. to replace all moments of nonzero CE order (except the
yet unknown terms $\Sigma_{\a\b}^{(1,2)}$) by suitable spatial
derivatives of the hydrodynamic variables. This is done via the
analysis of yet higher-order moments.

\subsection{Dynamics of higher-order moments, and closure}

By taking moments of the CE hierarchy, we derive the following
equations of motion:
\begin{itemize}
  \item Stress at order $\eps^{1}$:
    \begin{equation}
      \p_{t_1} \pi_{\a\b}^{(0)} + \p_{\g_1} \phi_{\a\b\g}^{(0)} = 
      \frac{1}{h} \left( \pi_{\a\b}^{*(1)} - \pi_{\a\b}^{(1)} \right) .
      \label{eq:2eps1}
    \end{equation}
  \item Third velocity moment at order $\eps^{1}$:
    \begin{equation}
      \p_{t_1}\phi_{\a\b\g}^{(0)} 
      + \p_{\delta_1} \psi_{\a\b\g\delta}^{(0)}
      = \frac{1}{h} \left( \phi_{\a\b\g}^{*(1)} - \phi_{\a\b\g}^{(1)} \right) .
      \label{eq:3eps1}
    \end{equation}
  \item Stress at order $\eps^{2}$:
    \begin{eqnarray}
      \nonumber
      & &
      \p_{t_2} \pi_{\a\b}^{(0)} + \frac{1}{2} \p_{t_1}
      \left( \pi_{\a\b}^{*(1)} + \pi_{\a\b}^{(1)} \right) 
      \\
      & &
      + \frac{1}{2} \p_{\g_1}
      \left( \phi_{\a\b\g}^{*(1)} + \phi_{\a\b\g}^{(1)} \right) =
      \frac{1}{h} \left( \pi_{\a\b}^{*(2)} - \pi_{\a\b}^{(2)} \right) ,
      \label{eq:2eps2Prelim}
    \end{eqnarray}
    or, again making use of the results of
    Sec. \ref{sec:SumVsDifference},
    \begin{eqnarray}
      \nonumber
      & &
      \p_{t_2} \pi_{\a\b}^{(0)} + \frac{1}{2} \frac{\g+1}{\g-1} \p_{t_1}
      \left( \pi_{\a\b}^{*(1)} - \pi_{\a\b}^{(1)} \right)
      \\
      \nonumber
      & &
      + \frac{1}{2} \frac{\g+1}{\g-1} \p_{\g_1}
      \left( \phi_{\a\b\g}^{*(1)} - \phi_{\a\b\g}^{(1)} \right) 
      \\
      \nonumber
      & &
      + \frac{1}{1 - \g} \p_{t_1} \Sigma_{\a\b}^{(1)}
      + \frac{1}{1 - \g} \p_{\g_1} \Xi_{\a\b\g}^{(1)}
      \\
      & &
      = \frac{1}{h} \left( \pi_{\a\b}^{*(2)} - \pi_{\a\b}^{(2)} \right) .
      \label{eq:2eps2}
    \end{eqnarray}
\end{itemize}
In principle, the strategy to proceed is as follows: Making use of
Eqs. \ref{eq:2eps1} and \ref{eq:3eps1}, we can express the terms
$\pi_{\a\b}^{*(1)} - \pi_{\a\b}^{(1)}$ and $\phi_{\a\b\g}^{*(1)} -
\phi_{\a\b\g}^{(1)}$ in Eq. \ref{eq:CE} via first-order derivatives of
zeroth-order moments. The latter, however, depend only on the
hydrodynamic variables $\rho$ and $\vect j$, for both of which we know
its respective first-order equation of motion (Eqs. \ref{eq:0eps1} and
\ref{eq:1eps1}). This ultimately allows us to express
$\pi_{\a\b}^{*(1)} - \pi_{\a\b}^{(1)}$ and $\phi_{\a\b\g}^{*(1)} -
\phi_{\a\b\g}^{(1)}$ in terms of spatial derivatives of hydrodynamic
variables, where, analogously to the case of the ideal gas, we neglect
terms of order $u^3$, i.~e.  assume that the flow velocity is
small. These results may then be inserted into Eq. \ref{eq:2eps2} as
well, to also eliminate $\pi_{\a\b}^{*(2)} -
\pi_{\a\b}^{(2)}$. Ultimately we then get a closed equation where,
except for the hydrodynamic variables, only the yet unknown moments of
the correction collision operator occur. These may then be adjusted in
order to make sure that all spurious terms cancel out. Most of the
details are done in App. \ref{app:hardwork}, and in practice we proceed
in a slightly different order.

As a first step, let us check that $\pi_{\a\b}^{*(1)} -
\pi_{\a\b}^{(1)}$ indeed corresponds to the Newtonian viscous
stress. As shown in App. \ref{app:hardwork}, one finds in the limit of
small $u$ (cf. Eq. \ref{eq:piminus}):
\begin{eqnarray}
\label{eq:piminusMainText}
& &
\frac{1}{h} \left (\pi_{\a\b}^{*(1)} - \pi_{\a\b}^{(1)} \right)
\\
\nonumber
& = &
\left( p - \rho \frac{\p p}{\p \rho} \right) \delta_{\a\b}
\p_{\g_1} u_\g + p \left( \p_{\b_1} u_\a + \p_{\a_1} u_\b \right)
\end{eqnarray}
or
\begin{eqnarray}
& &
\frac{\eps}{2} \frac{\g + 1}{\g -1}
\left (\pi_{\a\b}^{*(1)} - \pi_{\a\b}^{(1)} \right)
\\
\nonumber
& = &
\frac{h}{2} \frac{\g + 1}{\g -1}
\left[ \left( p - \rho \frac{\p p}{\p \rho} \right) \delta_{\a\b}
\p_{\g} u_\g + p \left( \p_{\a} u_\b + \p_{\b} u_\a \right) \right] ,
\end{eqnarray}
which is indeed the Newtonian viscous stress, however with shear ($\eta$)
and bulk ($\eta_V$) viscosities that depend on the state point:
\begin{eqnarray}
 \eta          & = & \frac{h}{2} \frac{1 + \g}{1 - \g} \, p , \\
 \eta_V / \eta & = & 
 1 + \frac{2}{d} - \frac{\rho}{p} \frac{\p p}{\p \rho} ,
\end{eqnarray}
where $d$ is the spatial dimension. It should be noted that the
condition $\eta_V > 0$ places a further restriction on the admissible
equation of state:
\begin{equation}
\frac{\p p}{\p \rho} < \left( 1 + \frac{2}{d} \right) \frac{p}{\rho} .
\end{equation}

This result may also be used directly to determine the correction
current $\wt{\vect j}$. From Eq. \ref{eq:wtj} and its discussion
we find
\begin{eqnarray}
  \label{eq:wtjClosed}
  \wt j_\a & = & - \frac{\eps^2 h^2}{12} \p_{\b_1}
  \frac{1}{h} \left( \pi_{\a\b}^{*(1)} - \pi_{\a\b}^{(1)} \right) 
  \\
  \nonumber
  & = &
  - \frac{h^2}{12} \p_{\b} \left[
  \left( p - \rho \frac{\p p}{\p \rho} \right) \delta_{\a\b}
  \p_{\g} u_\g + p \left( \p_{\b} u_\a + \p_{\a} u_\b \right)
  \right] .
\end{eqnarray}

We now use Eq. \ref{eq:2eps2} to eliminate the second-order
stress. Inserting this into Eq. \ref{eq:CE}, we find
\begin{align}
  & \p_t j_\a + \p_\b \pi_{\a\b}^{(0)}
  + \frac{\eps}{2} \frac{\g+1}{\g-1} \p_\b
  \left( \pi_{\a\b}^{*(1)} - \pi_{\a\b}^{(1)} \right) 
  = f_\a 
  \label{eq:CEf} \\
  & + \frac{\eps}{\g-1} \p_\b \Sigma_{\a\b}^{(1)} 
  + \frac{\eps^2 h}{2} \frac{\g+1}{(\g-1)^2}
  \p_\b \p_{t_1} \Sigma_{\a\b}^{(1)} 
  \notag\\ 
  & + \frac{\eps^2}{\g - 1} \p_\b \Sigma_{\a\b}^{(2)}
    + \frac{\eps h}{2} \frac{\g+1}{(\g-1)^2} 
      \p_\b \p_\g \Xi_{\a\b\g}^{(1)} 
  \notag \\
  & - \frac{\eps^2 h}{2} \frac{\g+1}{\g-1} \p_\b \p_{t_2} \pi_{\a\b}^{(0)}
  \notag \\
  & - \frac{\eps^2 h}{2}
  \frac{\g^2 + 4 \g + 1}{3 (\g-1)^2} \p_\b \p_{t_1} 
  \left( \pi_{\a\b}^{*(1)} - \pi_{\a\b}^{(1)} \right) 
  \notag \\
  & - \frac{\eps h}{2} \frac{\g^2 + 4 \g + 1}{3(\g-1)^2} \p_\b \p_\g
  \left( \phi_{\a\b\g}^{*(1)} - \phi_{\a\b\g}^{(1)} \right)  .
  \notag
\end{align} 

It is now easy to see that all the spurious terms can be eliminated by
proper construction of the second and third moments of the correction
collision operator (see Eqs. \ref{eq:SecondMomentDelta} and
\ref{eq:ThirdMomentDelta}), which have been up to now completely
arbitrary. By matching powers of $\eps$ and tensor ranks, one finds
that the following conditions have to be fulfilled:
\begin{equation}
 \label{eq:Sigma1Zero}
 \Sigma_{\a\b}^{(1)} = 0 ,
\end{equation}
\begin{equation}
  \Xi_{\a\b\g}^{(1)} = \frac{\g^2 + 4 \g + 1}{3(\g+1)} 
  \left( \phi_{\a\b\g}^{*(1)} - \phi_{\a\b\g}^{(1)} \right) ,
  \label{eq:xi1}
\end{equation}
\begin{eqnarray}
  \label{eq:sigma2}
  \Sigma_{\a\b}^{(2)} & = &
  h \frac{\g+1}{2} \, \p_{t_2} \pi_{\a\b}^{(0)} 
  \\
  \nonumber
  & + & 
  h \frac{\g^2 + 4 \g + 1}{6 (\g-1)} \,
  \p_{t_1} \left( \pi_{\a\b}^{*(1)} - \pi_{\a\b}^{(1)} \right) .
\end{eqnarray}

We proceed by further closing the equations, i.~e. by replacing the rh
sides of Eqs. \ref{eq:xi1} and \ref{eq:sigma2} with the corresponding
spatial derivatives of the hydrodynamic variables. This is done in
App. \ref{app:hardwork}, see Eqs. \ref{eq:phiminus},
\ref{eq:dt1piminus} and \ref{eq:dt2pi} (we do not repeat the lengthy
expressions derived there). The thus-derived derivatives then need to
be discretized in a consistent fashion, i.~e. correctly up to third
order in $\eps$. How to do this has already been discussed in
Sec. \ref{sec:centraldifference}.

\section{\label{sec:correction}
Correction collision operator}

The correction collision operator $\Delta_i^{corr}$ is still
unknown. However, we now know all the conditions that it has to
satisfy. Let us collect them here again:
\begin{itemize}
  \item Mass conservation:
    \begin{equation}
      \sum_i \Delta_i^{corr} = 0.
    \end{equation}
  \item Consistent momentum transfer:
    \begin{equation}
       \sum_i \Delta_i^{corr} c_{i\a} = (\gamma - 1) \tilde j_\a
    \end{equation}
    (see Eq. \ref{eq:CorrectionMomentumTransfer}), where
    $\wt{\vect j}$ is given by Eq. \ref{eq:wtjClosed}.
  \item Second moment:
     \begin{equation}
       \sum_{i} \Delta_i^{corr} c_{i\a} c_{i\b} = \Sigma_{\a\b}
       = \eps^2 \Sigma_{\a\b}^{(2)} ,
     \end{equation}
     where the rhs is given by Eqs. \ref{eq:sigma2},
     \ref{eq:dt1piminus} and \ref{eq:dt2pi}.
  \item Third moment:
     \begin{equation}
       \sum_{i} \Delta_i^{corr} c_{i\a} c_{i\b} c_{i\g} 
       = \Xi_{\a\b\g} = \eps \Xi_{\a\b\g}^{(1)} ,
     \end{equation}
     where the rhs is given by Eqs. \ref{eq:xi1} and \ref{eq:phiminus}.
\end{itemize}

A collision operator that yields these desired moments is
\begin{eqnarray}
  \label{eq:ConstructCollisionOperator}
  \Delta_i^{corr}
  & = & 
  (\g - 1) \, \frac{w_i}{c_s^2} \, \wt j_\a c_{i\a}
  \\
  \nonumber
  & + & \frac{w_i}{2 c_s^4} \, 
  \Sigma_{\a\b} \, (c_{i\a} c_{i\b} - c_s^2 \delta_{\a\b})
  \\ 
  \nonumber 
  & + & 
  \frac{w_i}{6 c_s^6} \,
  \Xi_{\a\b\g}^{\prime} \, (c_{i\a} c_{i\b} c_{i\g} 
  - c_s^2 \delta_{\a\b} c_{i\g} 
  \\
  \nonumber
  & - & c_s^2 \delta_{\a\g} c_{i\b} 
  - c_s^2 \delta_{\b\g} c_{i\a}) 
\end{eqnarray}
with
\begin{eqnarray}
& &
\Xi^{\prime}_{\a\b\g} 
\\
\nonumber
& = &
\Xi_{\a\b\g} - (\g - 1) c_s^2 \left(
\delta_{\a\b} \wt j_\g  + \delta_{\a\g} \wt j_\b + 
\delta_{\b\g} \wt j_\a \right) .
\end{eqnarray}
As in the case of the interface force operator, the weights $w_i$ and
the pressure $\rho c_s^2$ need to be evaluated at one fixed reference
density $\rho_0$, in order to avoid the occurence of additional
unwanted gradients.

\section{\label{sec:implicit} Implicit algorithm}

The correction current $\wt{\vect j}$ is given by Eq. \ref{eq:wtjClosed}.
However, the rhs of this equation depends on the streaming velocity
$\vect u = \vect j / \rho$, plus its gradients, while $\vect u$ in turn
depends on $\wt{\vect j}$, see Eq. \ref{eq:DefineJCurrent}. This
means that $\wt{\vect j}$ and $\vect u$ are defined \emph{implicitly}.
These observations suggest the following iterative procedure
to calculate the correction current:
\begin{enumerate}
\item On each lattice site, determine the density $\rho$.
\item On each lattice site, determine the force density
      $\vect f$, using the outlined finite-difference procedure.
\item On each lattice site, initialize $\wt{\vect j}$ by
      setting it to zero, or by taking the value from the
      previous time step.
\item On each lattice site, calculate $\vect j$ and $\vect u$
      from Eq. \ref{eq:DefineJCurrent}.
\item On each lattice site, calculate $\wt{\vect j}$ from Eq.
      \ref{eq:wtjClosed}, again using a finite-difference
      procedure.
\item Go to step 4, unless the iteration has converged.
\end{enumerate}
It should be noted that this problem is essentially a linear system of
equations, and hence it should in principle be amenable to more
sophisticated iterative solvers as well.

As soon as the iteration has converged, the values of all hydrodynamic
variables are available on all lattice sites. One may then proceed to
evaluate the equilibrium populations and the three contributions to
the collision operator. The collision is followed by a streaming step,
after which the procedure starts again. For a more detailed
description of the algorithm as a whole, see the Supplemental Material
\cite{zelkoduenweg_pre_supplemental_2014}.

\section{\label{sec:conclus}
Conclusions}

The present paper has dealt with the attempt to construct an
isothermal LB algorithm for gas-liquid coexistence, with the goal to
obtain a procedure that is (in the limit of sufficiently slow flows)
fully consistent with both hydrodynamics and thermodynamics. Motivated
by the success of LB methods for the ideal gas, we constructed the
method in close analogy to what is known from there. Central to our
approach is the observation that bulk and interfacial free energies
should enter the analysis at very different orders of the CE
expansion: The bulk free energy (or the bulk pressure) should be
encoded in the zeroth order, or the equilibrium populations, while the
interfacial force density, involving a third-order gradient, should
enter at third order. The present paper therefore directly builds upon
this observation, and constructs an algorithm that is systematically
shown to be consistent up to and including the third order, since this
is a \emph{necessary} condition for consistency of the method as a
whole. Up to now, to the best of our knowledge, has neither a CE
analysis of multiphase LB ever been done up to third order, nor has
any LB algorithm for a gas-liquid system been constructed that would
satisfy that consistency criterion. It is therefore hardly surprising
that so far multiphase LB methods have always been plagued by
artifacts like ``spurious currents''. Since the CE analysis as such
involves quite some tedious algebra, and the corresponding algorithm
needs substantial coding efforts, we have here confined ourselves to
the presentation of the theory only, while numerics is left for future
work.

The theoretical analysis has produced a wealth of interesting new
results, which we summarize here briefly. Most importantly, one needs
many velocity shells (our solution: D3Q59, D2Q21) in order to
accommodate all the isotropy constraints (CE consistency up to third
order requires isotropy of the weight moments up to sixth rank
tensors), plus the freedom to choose a non-trivial equation of
state. Since we wish to be able to define a thermodynamic entropy for
our system, we require the weights to be positive, which is only
possible if $p/\rho$ varies within narrow bounds. Furthermore, the
condition of positive bulk viscosity places yet another constraint on
the equation of state.  While the interface force density may be
determined fairly straightforwardly via a standard finite-difference
scheme, and the interface collision operator is constructed in direct
analogy to the coupling of LB to an external force density, a
completely new aspect is the occurence of a ``correction current''
$\wt{\vect j}$ that is necessary to ensure consistency of the
continuity equation up to third CE order. Unfortunately, one needs an
implicit (iterative) procedure to determine that current. In order to
systematically eliminate all spurious terms in the Navier-Stokes
equation, we finally construct a correction collision operator that
may be calculated from derivates of hydrodynamic variables via a
finite-difference scheme, which is a somewhat tedious though in
principle straightforward calculation.

While the newly developed algorithm has not yet been tested ---
neither in terms of efficiency, consistency, or accuracy, nor in terms
of its stability --- we believe that our theoretical results are
correct and interesting, and can form a solid basis for future
theoretical and numerical work in the field.

\begin{acknowledgments}
  This work was supported by the Volkswagen Foundation within the
  project ``Simulation Methods for Electrostatic and Hydrodynamic
  Interactions in Complex Systems'' (I/83 918). We thank Tony Ladd,
  Alexander Wagner, Julia Yeomans, Mike Cates, Ronojoy Adhikari,
  Fathollah Varnik, Jens Harting, Simone Melchionna, Taehun Lee and
  Ignacio Pagonabarraga for inspiring discussions.
\end{acknowledgments}

\appendix 

\section{\label{app:modelH} 
Interface force from Cahn-Hilliard free energy}

\subsection{Dissipation--free bulk hydrodynamics}

The continuity equation reads as
\begin{equation} \label{eq:Continuity}
\frac{\partial}{\partial t} \rho 
+ \partial_\alpha \left( \rho u_\alpha \right) = 0 .
\end{equation}
Introducing the convective derivative
\begin{equation} \label{eq:ConvDeriv}
\frac{D}{D t} = \frac{\partial}{\partial t} + u_\alpha \partial_\alpha ,
\end{equation}
this is rewritten as
\begin{equation} \label{eq:ContinuityConvDeriv}
\frac{D}{D t} \rho + \rho \partial_\alpha u_\alpha = 0 .
\end{equation}
From this, one can easily show the identity
\begin{equation} \label{eq:TransformConvDeriv}
\rho \frac{D}{D t} \left( \frac{\phi}{\rho} \right) =
\frac{\partial \phi}{\partial t} 
+ \partial_\alpha \left( u_\alpha \phi \right)
\end{equation}
for an arbitrary function $\phi$.
Introducing the momentum density $\vect j = \rho \vect u$, the Euler equation
in the presence of a force density $\vect f$ is written as
\begin{equation} \label{eq:Euler}
\frac{\partial}{\partial t} j_\alpha
+ \partial_\beta (u_\beta j_\alpha) + \partial_\alpha p = f_\alpha.
\end{equation}
Using Eq. \ref{eq:TransformConvDeriv}, this is rewritten as
\begin{equation}
\rho \frac{D}{D t} u_\alpha + \partial_\alpha p = f_\alpha .
\end{equation}
Therefore, we find for the kinetic energy, again using
Eq. \ref{eq:TransformConvDeriv}
\begin{eqnarray} \label{eq:MotionKineticEnergy}
&&
\frac{\partial}{\partial t} 
\left( \frac{1}{2} \rho u_\alpha u_\alpha \right) +
\partial_\beta \left( \frac{1}{2} \rho u_\beta u_\alpha u_\alpha \right) 
\\
\nonumber
& = &
\rho \frac{D}{D t} \left( \frac{1}{2} u_\alpha u_\alpha \right)
=
\rho u_\alpha \frac{D}{D t} u_\alpha
=
u_\alpha \left( f_\alpha - \partial_\alpha p \right) .
\end{eqnarray}

Let $e$ and $s$ denote the internal energy and entropy \emph{per unit mass},
respectively, such that internal energy and entropy density are given by
$\rho e$ and $\rho s$, respectively. If $E$ is the internal energy and $S$
the entropy, then the first law of thermodynamics for fixed particle number
$N$ or fixed total mass $M$ reads
\begin{equation}
dE = T dS - p dV;
\end{equation}
here $T$ is the temperature and $V$ the volume. Dividing
this equation by the total mass $M$, we obtain
\begin{equation}
de = T ds - p d \left( \frac{1}{\rho} \right)
   = T ds + \frac{p}{\rho^2} d\rho .
\end{equation}
Since we are studying dissipation--free hydrodynamics, there is
no entropy production, and the equation of motion for the entropy
is simply
\begin{equation}
\frac{D}{D t} s = 0.
\end{equation}
We therefore find
\begin{equation}
\rho \frac{D}{D t} e = \frac{p}{\rho} \frac{D}{D t} \rho =
- p \partial_\alpha u_\alpha .
\end{equation}
Again using Eq. \ref{eq:TransformConvDeriv}, we can rewrite this as
\begin{equation}
\frac{\partial}{\partial t} \left( \rho e \right)
+ \partial_\alpha \left( u_\alpha \rho e \right) =
- p \partial_\alpha u_\alpha .
\end{equation}
The equation of motion for the total energy density therefore
results to
\begin{eqnarray}
\nonumber
&&
\frac{\partial}{\partial t} 
\left( \frac{1}{2} \rho \vect u^2 + \rho e \right)
+ \partial_\alpha \left\{ u_\alpha 
\left( \frac{1}{2} \rho \vect u^2 + \rho e \right)
\right\} \\
& = &
u_\alpha \left( f_\alpha - \partial_\alpha p \right)
- p \partial_\alpha u_\alpha
\end{eqnarray}
or
\begin{eqnarray}
\nonumber
&& \frac{\partial}{\partial t} 
\left( \frac{1}{2} \rho \vect u^2 + \rho e \right)
+ \partial_\alpha \left\{ u_\alpha 
\left( \frac{1}{2} \rho \vect u^2 + \rho e + p \right)
\right\} \\
& = & u_\alpha f_\alpha .
\end{eqnarray}
For $\vect f = 0$, this is the conservation of energy. For a system
with periodic boundary conditions, we define the Hamiltonian in
$d$--dimensional space as
\begin{equation}
{\cal H} = \int d^d \vect r
\left( \frac{1}{2} \rho \vect u^2 + \rho e \right) ,
\end{equation}
and Gauss' theorem implies that it is only changed as a result
of the force density $\vect f$:
\begin{equation}
\frac{d}{dt} {\cal H} = \int d^d \vect r u_\alpha f_\alpha .
\end{equation}

\subsection{Inclusion of interfacial energies}

We now modify the Hamiltonian to also include an interfacial
term, i.~e.
\begin{equation}
{\cal H} = \int d^d \vect r
\left[ \frac{1}{2} \rho \vect u^2 + \rho e 
+ \frac{\kappa}{2} \left( \nabla \rho \right)^2
\right] ,
\end{equation}
with $\kappa > 0$. Furthermore, we require that the Hamiltonian
is conserved under the non--dissipative dynamics of the previous section,
i.~e. that the force term $\vect f$ is chosen in such a way that
\begin{equation}
\frac{d}{dt} {\cal H} = 0 .
\end{equation}
This however means
\begin{equation}
\int d^d \vect r u_\alpha f_\alpha +
\frac{\kappa}{2} \int d^d \vect r 
\frac{\partial}{\partial t} \left( \nabla \rho \right)^2 = 0 .
\end{equation}
Now, from the continuity equation we find
\begin{eqnarray}
\nonumber
&& \frac{1}{2} \frac{\partial}{\partial t} \left( \nabla \rho \right)^2
= \left( \partial_\beta \rho \right)
\partial_\beta \frac{\partial}{\partial t} \rho \\
& = & - \left( \partial_\beta \rho \right)
\partial_\beta \partial_\alpha \left( u_\alpha \rho \right) ,
\end{eqnarray}
resulting in
\begin{eqnarray}
\nonumber
\int d^d \vect r u_\alpha f_\alpha & = &
\kappa \int d^d \vect r \left( \partial_\beta \rho \right)
\partial_\beta \partial_\alpha \left( u_\alpha \rho \right) 
\\
& = &
\kappa \int d^d \vect r u_\alpha \rho
\partial_\alpha \partial_\beta \partial_\beta \rho ,
\end{eqnarray}
where in the last step we have done a two--fold partial
integration. Since this result must hold for any $\vect u$, we find for
the force density $\vect f$ the unique result
\begin{equation} \label{eq:ResultingForce2}
\vect f = \kappa \rho \nabla \nabla^2 \rho .
\end{equation}
The dynamics is therefore, by construction, energy--conserving. It is
however also momentum--conserving, since the total force applied
to the system vanishes:
\begin{equation}
\int d^d \vect r \vect f = 0 ;
\end{equation}
this latter relation is easily shown by inserting the
explicit formula Eq. \ref{eq:ResultingForce2}, and doing
a three--fold partial integration, which shows
\begin{equation}
\int d^d \vect r \vect f = - \int d^d \vect r \vect f .
\end{equation}
Alternatively, it is also possible to write $\vect f$ as the
divergence of a stress tensor. This latter approach has been mainly
pursued by Swift et al. \cite{swift_lattice_1995,swift_lattice_1996};
it has however the disadvantage that the stress tensor is in general
not unique, while the force term definitely is, as has been shown by
the present derivation.

\section{\label{app:EquationOfState} 
Equation of state}

The purpose of this appendix is to demonstrate, by explicit
construction, that it is possible to find an equation of state that
satisfies all the conditions for the present model.

We first recall that the equation of state is written as
\begin{equation}
  p(\rho) = \rho c_s^2 (\rho) ,
\end{equation}
and introduce the function
\begin{equation} 
  \label{eq:DefinePsiOfRho}
  \psi(\rho) := \frac{\partial}{\partial \rho} \ln c_s^2 (\rho) ,
\end{equation}
for which we note the identity
\begin{equation}
  \frac{\rho}{p} \frac{\partial p}{\partial \rho} =
  \rho \frac{\partial}{\partial \rho} \ln p =
  \rho \frac{\partial}{\partial \rho} \left( \ln \rho + \ln c_s^2 \right) =
  1 + \rho \psi(\rho) .
\end{equation}

Secondly, we recall all the conditions that the equation
of state has to satisfy:
\begin{itemize}
\item Positivity of weights:
      \begin{equation}
      \label{eq:PositivityOfWeights}
        c_{s, min}^2 < c_s^2 < c_{s, max}^2 ,
      \end{equation}
      where $c_{s, min}^2 = 0.3510760$, $c_{s, max}^2 = 1.333333$ in
      two dimensions, while $c_{s, min}^2 = 0.3850612$,
      $c_{s, max}^2 = 1.1917145$ in three dimensions (in lattice units,
      where lattice spacing and time step have been set to unity).
\item Two-phase coexistence: There must be some density interval
      for which the equation of state is unstable, i.~e.
      \begin{equation}
        \frac{\partial p}{\partial \rho} < 0 
      \end{equation}
      or
      \begin{equation}
        \label{eq:TwoPhaseCoexistence}
        \psi(\rho) < - \frac{1}{\rho} .
      \end{equation}
\item Positivity of the bulk viscosity (in $d$ dimensions):
      \begin{equation}
        \frac{\rho}{p} \frac{\partial p}{\partial \rho} < 1 + \frac{2}{d}
      \end{equation}
      or (for all values of $\rho$)
      \begin{equation}
        \label{eq:PositiveBulkViscosity}
        \psi(\rho) < \frac{2}{d} \frac{1}{\rho} .
      \end{equation}
\end{itemize}

Our strategy to find a valid equation of state therefore consists of
first constructing a function $\psi(\rho)$ that satisfies
Eqs. \ref{eq:TwoPhaseCoexistence} and \ref{eq:PositiveBulkViscosity}.
After picking some density value $\rho_0$ and its corresponding
$c_s^2$ value, we then find, via integration
\begin{equation}
  c_s^2 (\rho) = c_s^2 (\rho_0) \exp \left(
  \int_{\rho_0}^{\rho} d\rho' \psi(\rho') \right) ,
\end{equation}
and the final step is to verify that this function statisfies
Eq. \ref{eq:PositivityOfWeights} for all $\rho$ values.

In order to construct a very simple model function for $\psi$, we
choose two densities $\rho_1$ and $\rho_2$ with $\rho_1 < \rho_2$ and
set $\rho_3 = 2 \rho_2 - \rho_1$. Furthermore, we choose an amplitude
$A \ge 0$ and assume
\begin{equation}
  \psi(\rho) = \left\{
  \begin{array}{l l l}
    0 & \quad & \rho \le \rho_1 \\[1ex]
    A \sin \left( \pi \frac{\rho - \rho_1}{\rho_2 - \rho_1} \right)
    & \quad & \rho_1 < \rho < \rho_3 \\[1ex]
    0 & \quad & \rho_3 \le \rho .
  \end{array}
  \right.
\end{equation}

\begin{figure}
\begin{center}
  \includegraphics[width=0.45\textwidth]{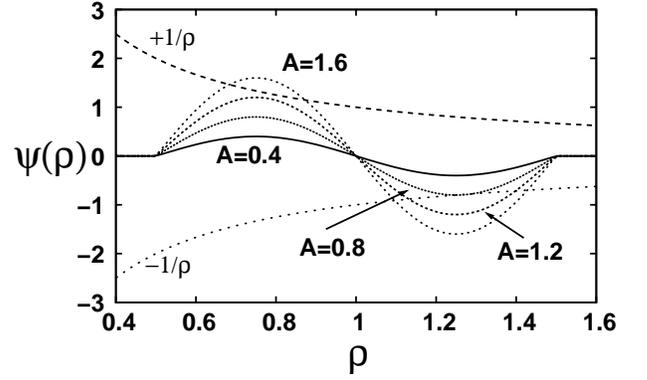}
\end{center}
\caption{Comparison of $\psi(\rho)$ with the right-hand sides
of Eqs. \ref{eq:TwoPhaseCoexistence} and \ref{eq:PositiveBulkViscosity},
for various amplitudes $A$.}
\label{fig:FirstConstruction}
\end{figure}

\begin{figure}
\begin{center}
  \includegraphics[width=0.45\textwidth]{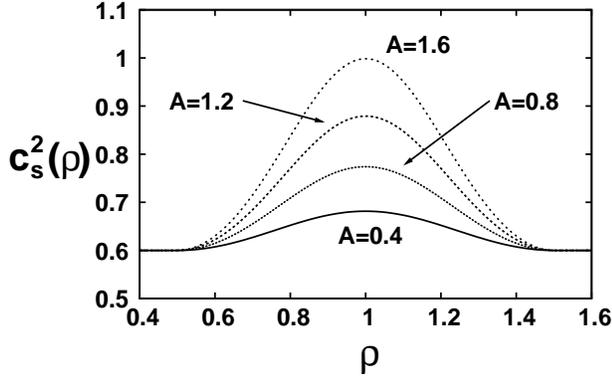}
\end{center}
\caption{$c_s^2 (\rho)$, for the parameters of Fig. \ref{fig:FirstConstruction},
and setting $c_s^2 (\rho_1) = 0.6$.}
\label{fig:cssquared}
\end{figure}

Figure \ref{fig:FirstConstruction} shows this function for various
values of the amplitude $A$, where we have chosen (in some arbitrary
units) $\rho_1= 0.5$, $\rho_2 = 1$. We also compare with the right
hand sides of Eqs. \ref{eq:TwoPhaseCoexistence} and
\ref{eq:PositiveBulkViscosity}, where we have picked the spatial
dimension $d = 2$. One sees that for large values of $A$ the condition
of positive bulk viscosity is violated, while for too small amplitudes
there is no two-phase coexistence. However, there is a certain window
of admissible amplitudes (for example, $A = 1.2$ in
Fig. \ref{fig:FirstConstruction}) where both conditions are met.  Now,
choosing $c_s^2(\rho_1) = 0.6$, we can also plot the function
$c_s^2(\rho)$ (Fig. \ref{fig:cssquared}), from which we see that
Eq. \ref{eq:PositivityOfWeights} is satisfied as well. It should be
noted that we have constructed our function $\psi$ in such a way that
\begin{equation}
  \int_{\rho_1}^{\rho_3} d\rho \psi(\rho) = 0 ,
\end{equation}
which means that $c_s^2$ takes the \emph{same} value for $\rho < \rho_1$
and $\rho > \rho_3$. This is a simplifying feature which is however
not necessary for the validity of the model.

\begin{figure}
\begin{center}
  \includegraphics[width=0.45\textwidth]{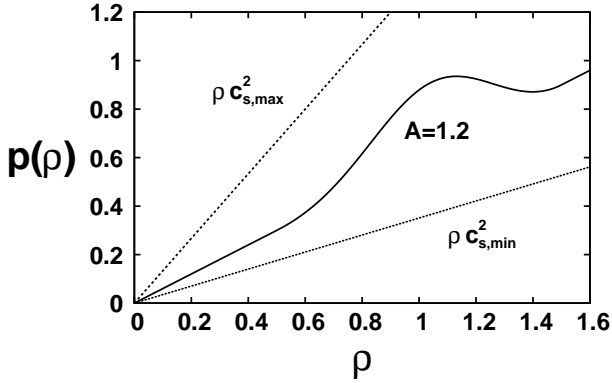}
\end{center}
\caption{$p (\rho)$, for the parameters of Fig. \ref{fig:cssquared},
setting $A = 1.2$. The limiting slopes that are given by Eq.
\ref{eq:PositivityOfWeights} are shown as well.}
\label{fig:PressureVsRho}
\end{figure}

\begin{figure}
\begin{center}
  \includegraphics[width=0.45\textwidth]{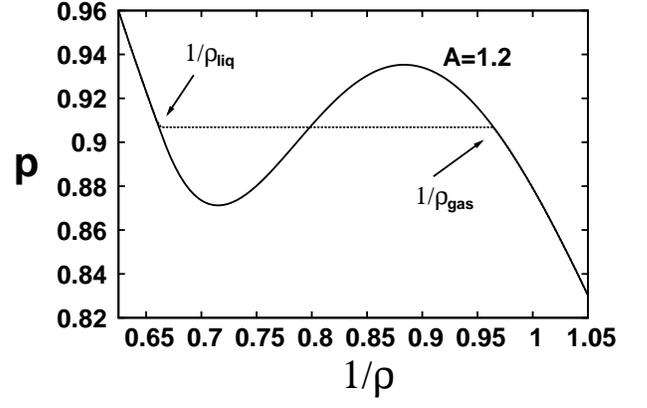}
\end{center}
\caption{Pressure $p$, for the parameters of Fig. \ref{fig:PressureVsRho},
as a function of specific volume $1/\rho$. Also shown is the Maxwell
construction that allows to determine the coexistence densities.}
\label{fig:PressureVsSpecVolume}
\end{figure}

Having thus found a valid function $c_s^2(\rho)$, we can now proceed
to look at the equation of state $p(\rho)$. This is done in
Fig. \ref{fig:PressureVsRho}, for the parameters of Figs.
\ref{fig:FirstConstruction} and \ref{fig:cssquared}, where we focus
attention on the valid amplitude $A = 1.2$. The same data are
re-plotted as a function of specific volume $1/\rho$ in Fig.
\ref{fig:PressureVsSpecVolume}, focusing on the interesting
coexistence region. This representation is amenable to the standard
Maxwell construction (also shown) that allows us to determine the
coexistence densities. The Maxwell construction was facilitated
by numerial root-finding, combined with a tabulated free energy
per unit mass $f$, which we found by numerically integrating
the relation $\partial f / \partial \rho = p / \rho^2$, and
normalizing by the requirement $f(\rho = \rho_1) = 0$.

It should be noted that the amplitude $A$ must be viewed as the
essential parameter that controls the thermodynamics of the
system. For $A = 0$, we recover the equation of state of an ideal
gas. For larger values of $A$, the equation of state more and more
deviates from ideality, until we reach a value beyond which
the equation $\psi(\rho) = - 1 / \rho$ has solutions. This is
the system's (Mean Field) critical point. From then on, the
equation of state assumes a more and more pronounced unstable region
($\partial p / \partial \rho < 0$), indicative of two-phase
coexistence, until finally $A$ becomes so large that the bulk
viscosity becomes negative --- this corresponds to a situation where
the system has been quenched so deeply into the two-phase region that
the LB model with its limited set of velocities is no longer able to
represent the physics in a consistent and numerically stable fashion.

\begin{figure}
\begin{center}
  \includegraphics[width=0.45\textwidth]{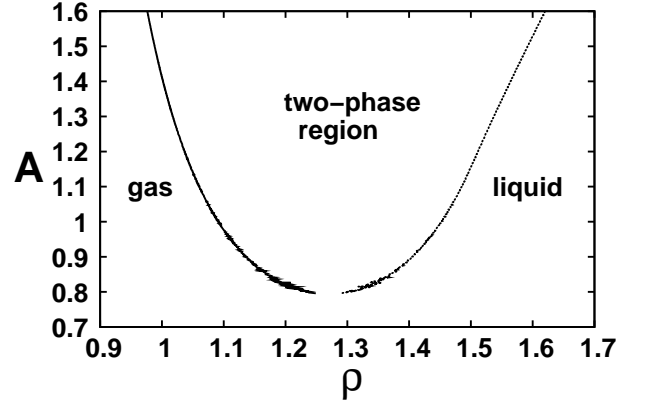}
\end{center}
\caption{Phase diagram of the model in the $\rho$ vs. $A$ plane.}
\label{fig:BulkPhaseDiagram}
\end{figure}

By systematically solving the Maxwell construction for various
values of $A$, we can finally find the system's bulk
phase diagram in the $\rho$ vs. $A$ plane. It is presented
in Fig.~\ref{fig:BulkPhaseDiagram}.

\section{\label{app:hardwork}
Miscellaneous expressions in the
derivation of the closure}

\subsection{$\pi_{\a\b}^{*(1)} - \pi_{\a\b}^{(1)}$}

$\pi_{\a\b}^{*(1)} - \pi_{\a\b}^{(1)}$ is determined via Eq. \ref{eq:2eps1}:
\begin{equation}
  \frac{1}{h} \left (\pi_{\a\b}^{*(1)} - \pi_{\a\b}^{(1)} \right) = 
  \p_{t_1} \pi_{\a\b}^{(0)} + \p_{\g_1} \phi_{\a\b\g}^{(0)} .
\end{equation}
It should be noted that $\pi_{\a\b}^{(0)}$ and $\phi_{\a\b\g}^{(0)}$
are functions only of the hydrodynamic variables (see
Eqs. \ref{eq:euler00} and \ref{eq:phi00}). Furthermore, at the
first-order level of the CE expansion, the dynamics of the
hydrodynamic variables is simply given by the continuity and the Euler
equations (see Eqs. \ref{eq:0eps1} and \ref{eq:1eps1}). Now we observe
that, neglecting terms of order $u^3$, we can write
\begin{equation}
\phi_{\a\b\g}^{(0)} = \pi_{\a\b}^{(0)} u_\g + 
p \left( u_\a \delta_{\b\g} + u_\b \delta_{\a\g} \right) ,
\end{equation}
i.~e.
\begin{equation}
\p_{\g_1} \phi_{\a\b\g}^{(0)} = \p_{\g_1} \left( \pi_{\a\b}^{(0)} u_\g \right)
+ \p_{\a_1} (p u_\b) + \p_{\b_1} (p u_\a) .
\end{equation}
Therefore we may write (cf. also App. \ref{app:modelH}):
\begin{eqnarray}
\label{eq:piminus}
& &
\frac{1}{h} \left (\pi_{\a\b}^{*(1)} - \pi_{\a\b}^{(1)} \right)
\\
\nonumber
& = &
\rho \frac{D}{D t_1} \left( \frac{1}{\rho} \pi^{(0)}_{\a\b} \right)
+ \p_{\a_1} (p u_\b) + \p_{\b_1} (p u_\a)
\\
\nonumber
& = &
\rho \frac{D}{D t_1} \left( \frac{p}{\rho} \delta_{\a\b} + u_\a u_\b \right)
+ \p_{\a_1} (p u_\b) + \p_{\b_1} (p u_\a)
\\
\nonumber
& = &
- \rho^2 \frac{\p}{\p \rho} \left( \frac{p}{\rho} \right) \delta_{\a\b}
\p_{\g_1} u_\g  - u_\b \p_{\a_1} p - u_\a \p_{\b_1} p 
\\
\nonumber
& + & \p_{\a_1} (p u_\b) + \p_{\b_1} (p u_\a)
\\
\nonumber
& = &
\left( p - \rho \frac{\p p}{\p \rho} \right) \delta_{\a\b}
\p_{\g_1} u_\g + p \left( \p_{\a_1} u_\b + \p_{\b_1} u_\a \right) .
\end{eqnarray}

\subsection{$\phi_{\a\b\g}^{*(1)} - \phi_{\a\b\g}^{(1)}$}

This expression is derived from Eq. \ref{eq:3eps1},
\begin{equation}
  \frac{1}{h} \left (\phi_{\a\b\g}^{*(1)} - \phi_{\a\b\g}^{(1)} \right)
  = \p_{t_1} \phi_{\a\b\g}^{(0)} + \p_{\delta_1} \psi_{\a\b\g\delta}^{(0)} ,
\end{equation}
and we proceed in a quite analogous fashion. First, from Eqs.
\ref{eq:euler00} -- \ref{eq:psi00} we conclude
\begin{equation}
\psi^{(0)}_{\a\b\g\delta} = \phi^{(0)}_{\a\b\g} u_\delta
+ \frac{p}{\rho} \left( \pi^{(0)}_{\a\b} \delta_{\g\delta}
+ \pi^{(0)}_{\a\g} \delta_{\b\delta}
+ \pi^{(0)}_{\b\g} \delta_{\a\delta} \right)
\end{equation}
and
\begin{eqnarray}
\p_{\delta_1} \psi^{(0)}_{\a\b\g\delta}
& = & \p_{\delta_1} \left( \phi^{(0)}_{\a\b\g} u_\delta \right)
  +   \p_{\g_1} \left( \frac{p}{\rho} \pi^{(0)}_{\a\b} \right)
\\
\nonumber
& + & \p_{\b_1} \left( \frac{p}{\rho} \pi^{(0)}_{\a\g} \right)
  +   \p_{\a_1} \left( \frac{p}{\rho} \pi^{(0)}_{\b\g} \right) .
\end{eqnarray}
From this we conclude (note that again $\p_{t_1}$, $\p_{\delta_1}$
implies simple Euler dynamics)
\begin{eqnarray}
& &
\frac{1}{h} \left (\phi_{\a\b\g}^{*(1)} - \phi_{\a\b\g}^{(1)} \right)
\\
\nonumber
& = &
\rho \frac{D}{D t_1} \left( \frac{1}{\rho} \phi^{(0)}_{\a\b\g} \right)
+ \p_{\g_1} \left( \frac{p}{\rho} \pi^{(0)}_{\a\b} \right) 
\\
\nonumber
& + & \p_{\b_1} \left( \frac{p}{\rho} \pi^{(0)}_{\a\g} \right)
  +   \p_{\a_1} \left( \frac{p}{\rho} \pi^{(0)}_{\b\g} \right)
\\
\nonumber
& = &
\rho \delta_{\a\b} \frac{D}{D t_1} \left( \frac{p}{\rho} u_\g \right)
+ \p_{\g_1} \left( \frac{p}{\rho} \pi^{(0)}_{\a\b} \right) 
\\
\nonumber
& + &
\rho \delta_{\a\g} \frac{D}{D t_1} \left( \frac{p}{\rho} u_\b \right)
+ \p_{\b_1} \left( \frac{p}{\rho} \pi^{(0)}_{\a\g} \right)
\\
\nonumber
& + &
\rho \delta_{\b\g} \frac{D}{D t_1} \left( \frac{p}{\rho} u_\a \right)
+ \p_{\a_1} \left( \frac{p}{\rho} \pi^{(0)}_{\b\g} \right)
\\
\nonumber
& \equiv &
\rho \delta_{\a\b} \frac{D}{D t_1} \left( \frac{p}{\rho} u_\g \right)
+ \p_{\g_1} \left( \frac{p}{\rho} \pi^{(0)}_{\a\b} \right) 
+ \text{perm.}
\end{eqnarray}
Now,
\begin{eqnarray}
& & 
\rho \frac{D}{D t_1} \left( \frac{p}{\rho} u_\g \right)
\\
\nonumber
& = &
- \frac{p}{\rho} \p_{\g_1} p
- \rho^2 \frac{\p}{\p \rho} \left( \frac{p}{\rho} \right)
u_\g \p_{\delta_1} u_\delta
\\
\nonumber
& = &
- \p_{\g_1} \left( \frac{p^2}{\rho} \right)
+ p \p_{\g_1} \left( \frac{p}{\rho} \right)
- \rho^2 \frac{\p}{\p \rho} \left( \frac{p}{\rho} \right)
u_\g \p_{\delta_1} u_\delta
\end{eqnarray}
and
\begin{equation}
\p_{\g_1} \left( \frac{p}{\rho} \pi^{(0)}_{\a\b} \right)
= \delta_{\a\b} \p_{\g_1} \left( \frac{p^2}{\rho} \right)
+ \p_{\g_1} \left( p u_\a u_\b \right) .
\end{equation}
Hence,
\begin{eqnarray}
\label{eq:phiminus}
& &
\frac{1}{h} \left (\phi_{\a\b\g}^{*(1)} - \phi_{\a\b\g}^{(1)} \right)
\\
\nonumber
& = &
\p_{\g_1} \left( p u_\a u_\b \right) 
\\
\nonumber
& + &
\delta_{\a\b} \left[ p \p_{\g_1} \left( \frac{p}{\rho} \right)
- \rho^2 \frac{\p}{\p \rho} \left( \frac{p}{\rho} \right)
u_\g \p_{\delta_1} u_\delta \right]
+ \text{perm.}
\end{eqnarray}

\subsection{$\p_{t_1}  ( \pi_{\a\b}^{*(1)} - \pi_{\a\b}^{(1)})$}

From Eq. \ref{eq:piminus} we find

\begin{eqnarray}
  & &
  \frac{1}{h} \p_{t_1} \left( \pi_{\a\b}^{*(1)} - \pi_{\a\b}^{(1)} \right)
  \\
  \nonumber
  & = &
  \p_{t_1} \left[
  p \left( \p_{\a_1} u_\b + \p_{\b_1} u_\a \right)
  + 
  \left( p - \frac{\p p}{\p \rho} \rho \right) \delta_{\a\b} \p_{\g_1} u_\g 
  \right] 
  \\
  \nonumber
  & = &
  \frac{\p p}{\p \rho} (\p_{t_1} \rho) (\p_{\a_1} u_\b + \p_{\b_1} u_\a) 
  \\
  \nonumber
  & + &
  p \p_{\a_1} \p_{t_1} u_\b + p \p_{\b_1} \p_{t_1} u_\a 
  \\
  \nonumber
  & - &
  \frac{\p^2 p}{\p \rho^2} \rho
  (\p_{t_1} \rho ) \delta_{\a\b} \p_{\g_1} u_\g 
  \\
  \nonumber
  & + &
  \left( p - \frac{\p p}{\p \rho}\rho \right) \delta_{\a\b}
  \p_{\g_1} \p_{t_1} u_\g .
\end{eqnarray}
Again, we have simple Euler dynamics and we can write
\begin{equation}
\frac{D}{D t_1} u_\a = \left( \p_{t_1} + u_\g \p_{\g_1} \right) u_\a
= - \frac{1}{\rho} \p_{\a_1} p
\end{equation}
or
\begin{equation}
\p_{t_1} u_\a = - \frac{1}{\rho} \p_{\a_1} p - u_\g \p_{\g_1} u_\a .
\end{equation}
Furthermore, we have
\begin{equation}
\p_{t_1} \rho = - \p_{\g_1} j_\g .
\end{equation}
Inserting these results, we find
\begin{eqnarray}
  \label{eq:dt1piminus}
  & &
  \frac{1}{h} \p_{t_1} \left( \pi_{\a\b}^{*(1)} - \pi_{\a\b}^{(1)} \right)
  \\
  \nonumber
  & = &
  - \frac{\p p}{\p \rho} (\p_{\g_1} j_\g) (\p_{\a_1} u_\b + \p_{\b_1} u_\a)
  \\
  \nonumber
  & - &
  p \p_{\a_1} \left[ \frac{1}{\rho} \p_{\b_1} p + u_\g \p_{\g_1} u_\b \right]
  \\
  \nonumber
  & - &
  p \p_{\b_1} \left[ \frac{1}{\rho} \p_{\a_1} p + u_\g \p_{\g_1} u_\a \right]
  \\
  \nonumber
  & - &
  \left(p - \frac{\p p}{\p \rho} \rho \right) \delta_{\a\b}
  \p_{\g_1} \left( \frac{1}{\rho} \p_{\g_1} p + 
  u_\delta \p_{\delta_1} u_\g \right)
  \\
  \nonumber
  & + &
  \frac{\p^2 p}{\p \rho^2} \rho  \delta_{\a\b} 
  (\p_{\g_1} u_\g) (\p_{\delta_1} j_\delta) .
\end{eqnarray}

\subsection{$\p_{t_2}  \pi_{\a\b}^{(0)}$}

On the $t_2$ time scale the dynamics is simply given by
(cf. Eqs. \ref{eq:0eps2} and \ref{eq:1eps2}, taking into account that
$\wt{\vect j}^{(1)}$ vanishes)

\begin{eqnarray}
  \p_{t_2} \rho  & = & 0 ,
  \\
  \p_{t_2} j_\a & = & - \frac{1}{2} \p_{\b_1} 
  \left( \pi_{\a\b}^{*(1)} + \pi_{\a\b}^{(1)} \right)
  \\
  \nonumber
  & = &
  - \frac{1}{2} \frac{\g+1}{\g-1} \p_{\b_1}
  \left( \pi_{\a\b}^{*(1)} - \pi_{\a\b}^{(1)} \right)
  \\
  \nonumber
  & = &
  - \frac{h}{2} \frac{\g+1}{\g-1} \p_{\a_1}
  \left[
  \left( p - \rho \frac{\p p}{\p \rho} \right)
  \p_{\g_1} u_\g
  \right]
  \\
  \nonumber
  & &
  - \frac{h}{2} \frac{\g+1}{\g-1} \p_{\b_1}
  \left[
  p \left( \p_{\a_1} u_\b + \p_{\b_1} u_\a \right)
  \right] ,
\end{eqnarray}
where in the transformations of $\p_{t_2} j_\a$ we have
made use of the results of Sec. \ref{sec:SumVsDifference},
taking into account that $\Sigma^{(1)}_{\a\b}$ vanishes,
plus of Eq. \ref{eq:piminus}. Therefore,
\begin{eqnarray}
  \p_{t_2} u_\a
  & = &
  - \frac{h}{2} \frac{\g+1}{\g-1} \frac{1}{\rho} \p_{\a_1}
  \left[
  \left( p - \rho \frac{\p p}{\p \rho} \right)
  \p_{\g_1} u_\g
  \right]
  \\
  \nonumber
  & &
  - \frac{h}{2} \frac{\g+1}{\g-1} \frac{1}{\rho} \p_{\b_1}
  \left[
  p \left( \p_{\a_1} u_\b + \p_{\b_1} u_\a \right)
  \right] .
\end{eqnarray}
From this, we conclude
\begin{eqnarray}
\label{eq:dt2pi}
\p_{t_2} \pi^{(0)}_{\a\b} & = &
\p_{t_2} \left( p \delta_{\a\b} + \rho u_\a u_\b \right)
\\
\nonumber
& = &
\rho \left( u_\a \p_{t_2} u_\b + u_\b \p_{t_2} u_\a \right)
\\
\nonumber
& = &
  - \frac{h}{2} \frac{\g+1}{\g-1} u_\a \p_{\b_1}
  \left[
  \left( p - \rho \frac{\p p}{\p \rho} \right)
  \p_{\g_1} u_\g
  \right]
\\
\nonumber
& &
  - \frac{h}{2} \frac{\g+1}{\g-1} u_\b \p_{\a_1}
  \left[
  \left( p - \rho \frac{\p p}{\p \rho} \right)
  \p_{\g_1} u_\g
  \right]
\\
\nonumber
& &
  - \frac{h}{2} \frac{\g+1}{\g-1} u_\b \p_{\g_1}
  \left[
  p \left( \p_{\a_1} u_\g + \p_{\g_1} u_\a \right)
  \right]
\\
\nonumber
& &
  - \frac{h}{2} \frac{\g+1}{\g-1} u_\a \p_{\g_1}
  \left[
  p \left( \p_{\b_1} u_\g + \p_{\g_1} u_\b \right)
  \right] .
\end{eqnarray}
%


%

\end{document}


\title{Supplemental material to: Consistent two-phase Lattice Boltzmann
model for gas-liquid systems}

\author{Jasna Zelko}
\affiliation{Max Planck Institute for Polymer Research,
Ackermannweg 10, 55128 Mainz, Germany}
\author{Burkhard D\"unweg}
\affiliation{Max Planck Institute for Polymer Research,
Ackermannweg 10, 55128 Mainz, Germany}
\affiliation{Condensed Matter Physics, TU Darmstadt,
Karolinenplatz 5, 64289 Darmstadt, Germany}
\affiliation{Department of Chemical Engineering,
Monash University, Clayton, Victoria 3800, Australia}

\date{\today}

\begin{abstract}
  We here provide a step-by-step summary of the algorithm
  derived in the main text.
\end{abstract}

\maketitle

We assume that the meaning of symbols is sufficiently clear from the
main text; we do not repeat their definitions here. It should be noted
that the algorithm implies the approximate evaluation of lots of
spatial derivatives by finite-difference operators. By pulling the
differential operators further to the left or to the right, one can
thus generate a lot of different algorithms, which all should give the
same asymptotic result (i.~e. should all be consistent with the
Navier-Stokes equation within the desired third-order Chapman-Enskog
accuracy), but which are numerically different. We are far from sure
if the formulation we have chosen is the most accurate or the most
efficient or the most convenient one. Most likely it is none of these;
it is simply one possible fairly arbitrary formulation out of many,
which has resulted from our analytical formulas. Now, the algorithm
proceeds via the steps given below. Please note that every step
starting with ``at each lattice site'' means a separate sweep through
the entire lattice, which must be completed before the next step may be
done.

\begin{enumerate}

\item At each lattice site $\vect r$: Set the correction
      current at that site to zero:
      \begin{equation}
        \tilde{\vect{j}} (\vect r) = 0 .
      \end{equation}
      At the same site: Evaluate the zeroth and first
      velocity moments:
      \begin{eqnarray}
        \rho & = & \sum_i n_i , \\
        \vect{j}_0 & = & \sum_i n_i \vect{c}_i .
      \end{eqnarray}
      Furthermore, evaluate (at the same lattice site)
      the equation of state to obtain for later use:
      \begin{itemize}
        \item the pressure $p$;
        \item $\partial p / \partial \rho$;
        \item $p - \rho (\partial p / \partial \rho) 
              = - \rho^2 (\partial / \partial \rho) (p / \rho)$;
        \item $\rho \partial^2 p / \partial \rho^2$.
      \end{itemize}

\item At each lattice site, evaluate the local interface force
      \begin{equation}
        f_{\alpha} = \kappa \rho
        \partial_\alpha \partial_\beta \partial_\beta \rho ,
      \end{equation}
      making use of the discretized differential operator
      derived in Sec. IV E. Calculate (at the same site)
      \begin{equation}
        \vect{j}_0^\prime = \vect{j}_0 + \frac{h}{2} \vect{f} .
      \end{equation}

\item At each lattice site, calculate the flow velocity
      \begin{equation}
        \vect u = \rho^{-1} \left( \vect{j}_0^\prime
        + \tilde{\vect{j}} \right) .
      \end{equation}

\item At each lattice site, calculate the tensor of derivatives
      $\partial_\alpha u_\beta$, using the discretized differential
      operator derived in Sec. IV B, and from this the tensor
      \begin{eqnarray}
        &&
        q_{\a \b} = 
        \\
        \nonumber
        &&
        \left( p - \rho \frac{\p p}{\p \rho} \right) \delta_{\a\b}
        \p_{\g} u_\g + p \left( \p_{\b} u_\a + \p_{\a} u_\b \right) .
      \end{eqnarray}

\item At each lattice site, evaluate
      \begin{equation}
        \tilde j_\alpha^{new} = - \frac{h^2}{12} \p_{\b} q_{\a \b} ,
      \end{equation}
      again using the first-order differential operator
      from Sec. IV B.

\item Check if the fields $\tilde{\vect{j}}^{new} (\vect r)$ and
      $\tilde{\vect{j}} (\vect r)$ are identical, within the
      convergence criterion of the iteration.

\item If no: Set $\tilde{\vect{j}} = \tilde{\vect{j}}^{new}$ on each
      lattice site; go to step 3.

\item If yes: Set $\tilde{\vect{j}} = \tilde{\vect{j}}^{new}$ on each
      lattice site; calculate on the same site the final value
      for the momentum density
      \begin{equation}
        \vect j = \vect{j}_0^\prime + \tilde{\vect{j}}
      \end{equation}
      and the flow velocity
      \begin{equation}
        \vect u = \rho^{-1} \vect{j} ,
      \end{equation}
      and continue.

\item At this point, the fields $\rho$, $p$, $\vect f$, $\vect j$
      and $\vect u$, the correction current $\tilde{\vect{j}}$,
      as well as various expressions involving
      thermodynamic derivatives of $p$,
      are available at every lattice site. These data are then
      used to evaluate various expressions needed for the
      correction collision operator, by making use of finite
      differences.

\item At every lattice site: Evaluate the 1st order lattice
      derivatives
      \begin{itemize}
        \item $\p_\a u_\b$
        \item $\p_\a j_\b$
        \item $\p_\a p$
        \item $\p_\a \left( p u_\b u_\g \right)$
        \item $\p_\a (p / \rho)$
      \end{itemize}

\item At every lattice site: Evaluate the 1st order
      lattice derivatives
      \begin{itemize}
        \item $\p_\a \left[ 
              \left(p - \rho \partial p / \partial \rho \right)
              \partial_\g u_\g \right]$
        \item $\p_\g \left[ p
              \left( \p_\a u_\b + \p_\b u_\a \right) \right]$
        \item $\p_\a \left[ \rho^{-1} \p_\b p + u_\g \p_\g u_\b
              \right]$
      \end{itemize}
      and from these the tensors
      \begin{eqnarray}
        &&
        \Sigma_{\a\b} =
        \\
        \nonumber
        &&
        - \frac{h^2}{4} \frac{(\g + 1)^2}{\g - 1} \Big\{
        \\
        \nonumber
        &&
        u_\a \p_{\b} \left[
        \left( p - \rho \frac{\p p}{\p \rho} \right)
        \p_{\g} u_\g \right]
        \\
        \nonumber
        & + &
        u_\b \p_{\a} \left[
        \left( p - \rho \frac{\p p}{\p \rho} \right)
        \p_{\g} u_\g \right]
        \\
        \nonumber
        & + &
        u_\b \p_{\g} \left[
        p \left( \p_{\a} u_\g + \p_{\g} u_\a \right) \right]
        \\
        \nonumber
        & + &
        u_\a \p_{\g} \left[
        p \left( \p_{\b} u_\g + \p_{\g} u_\b \right) \right]
        \Big\}
        \\
        \nonumber    
        & + &
        \frac{h^2}{6} \frac{\g^2 + 4 \g + 1}{\g - 1} \Big\{
        \\
        \nonumber
        & - &
        \frac{\p p}{\p \rho} (\p_{\g} j_\g) (\p_{\a} u_\b + \p_{\b} u_\a)
        \\
        \nonumber
        & - &
        p \p_{\a} \left[ \frac{1}{\rho} \p_{\b} p + u_\g \p_{\g} u_\b \right]
        \\
        \nonumber
        & - &
        p \p_{\b} \left[ \frac{1}{\rho} \p_{\a} p + u_\g \p_{\g} u_\a \right]
        \\
        \nonumber
        & - &
        \left(p - \frac{\p p}{\p \rho} \rho \right) \delta_{\a\b}
        \p_{\g} \left( \frac{1}{\rho} \p_{\g} p + 
        u_\delta \p_{\delta} u_\g \right)
        \\
        \nonumber
        & + &
        \frac{\p^2 p}{\p \rho^2} \rho  \delta_{\a\b} 
        (\p_{\g} u_\g) (\p_{\delta} j_\delta) \Big\}
      \end{eqnarray}
      and
      \begin{eqnarray}
        &&
        \Xi^\prime_{\a \b \g} =
        \\
        \nonumber
        &&
        \frac{h}{3} \frac{\g^2 + 4 \g + 1}{\g + 1} \Big\{
        \p_{\g} \left( p u_\a u_\b \right) +
        \\
        \nonumber
        &&
        \delta_{\a\b} \left[ p \p_{\g} \left( \frac{p}{\rho} \right)
        - \rho^2 \frac{\p}{\p \rho} \left( \frac{p}{\rho} \right)
        u_\g \p_{\delta} u_\delta \right]
        \\
        \nonumber
        &&
        + \text{perm.} 
        \Big\}
        \\
        \nonumber
        &&
        + (1 - \g) (p / \rho) \left(
        \delta_{\a\b} \wt j_\g  + \delta_{\a\g} \wt j_\b + 
        \delta_{\b\g} \wt j_\a \right) .
      \end{eqnarray}

      Then, at this lattice site, evaluate for each velocity $i$
      \begin{itemize}
        \item the weight $w_i$ for the density at that site
              (see Sec. III);
        \item the equilibrium population (note $c_s^2 = p/\rho$)
              \begin{eqnarray}
                &&
                \quad \quad 
                n_i^{eq} =
                \\
                \nonumber
                && 
                \quad \quad
                w_i \rho 
                \left( 1 + \frac{\vect u \cdot \vect c_i}{c_s^2} 
                + \frac{(\vect u \cdot \vect c_i)^2}{2 c_s^4} -
                \frac{u^2}{2 c_s^2} \right) ;
              \end{eqnarray}
        \item the bulk collision operator
              \begin{equation}
                \Delta_i^{bulk} = (\gamma - 1) (n_i - n_i^{eq}) ;
              \end{equation}
        \item the interface collision operator
              \begin{equation}
                \Delta_i^{int} = \frac{1 + \g}{2} h 
                  \frac{w_i(\rho_0)}{c_s^2(\rho_0)} c_{i\a} f_{\a} ,
              \end{equation}
              where $\rho_0$ is some arbitrary fixed density;
        \item and the correction collision operator
              \begin{eqnarray}
                &&
                \Delta_i^{corr} =
                \\
                \nonumber
                &&
                (\g - 1) \, \frac{w_i}{c_s^2} \, \wt j_\a c_{i\a}
                \\
                \nonumber
                &&
                + \frac{w_i}{2 c_s^4} \, 
                \Sigma_{\a\b} \, (c_{i\a} c_{i\b} - c_s^2 \delta_{\a\b})
                \\
                \nonumber 
                && 
                + \frac{w_i}{6 c_s^6} \,
                \Xi_{\a\b\g}^{\prime} \, (c_{i\a} c_{i\b} c_{i\g} 
                - c_s^2 \delta_{\a\b} c_{i\g} 
                \\
                \nonumber
                && 
                - c_s^2 \delta_{\a\g} c_{i\b} 
                - c_s^2 \delta_{\b\g} c_{i\a}) ,
              \end{eqnarray}
      \end{itemize}
      where again $w_i$ and $c_s^2$ are evaluated at some fixed
      density $\rho_0$.

      Then, for that velocity do the actual Lattice Boltzmann
      update, and store the output on a new lattice to avoid
      possible overwriting of data:
      \begin{eqnarray}
        n_i^{new} (\vect r + \vect c_i h) & = & n_i (\vect r) \\
        \nonumber
        & + & \Delta_i^{bulk} + \Delta_i^{int} + \Delta_i^{corr} .
      \end{eqnarray}

\item After all the collision and streaming steps are done:
      Rename $n_i^{new}$ to $n_i$ and increment time by $h$;
      go to step 1.

\end{enumerate}